\documentclass[10pt,prd,aps,twocolumn,nofootinbib,superscriptaddress,showpacs,floatfix,amssymb,amsmath]{revtex4-1}
\usepackage{slashed}
\usepackage{epsfig}
\usepackage{multirow}

\usepackage{bbold}
\usepackage{float}
\usepackage{hhline}
\usepackage[table]{xcolor}
\usepackage{array,multirow}
\usepackage{simplewick}

\usepackage{xcolor}
\usepackage{ulem}
\definecolor{purple}{rgb}{0.8,0,0.6}

\usepackage{graphicx}
\usepackage{bm}
\usepackage{combelow}
\usepackage{braket}

\newcommand{\MF}{{\mathrm{MF}}}
\newcommand{\M}{LSM${}_q$}
\newcommand{\beqs}{\begin{subequations}}
\newcommand{\eeqs}{\end{subequations}\\[-2mm]\noindent}
\newcommand{\beqn}{\begin{eqnarray}}
\newcommand{\eeqn}{\end{eqnarray}}
\newcommand{\eq}[1]{(\ref{#1})}

\newcommand{\cL}{{\cal L}}

\newcommand{\cM}{{\cal M}}

\newcommand{\MeV}{\,{\mathrm{MeV}}}

\newcommand{\ZP}{{\mathrm{ZP}}}

\newcommand{\Z}{{\mathbb{Z}}}

\newcommand{\bs}{\boldsymbol}
\newcommand{\p}{{\bs p}}
\newcommand{\x}{{\bs x}}
\newcommand{\h}{{\mathbb h}}
\newcommand{\avr}[1]{{\left\langle #1 \right\rangle}}

\usepackage{color}
\definecolor{purple}{rgb}{0.8,0,0.6}

\begin{document}

\title{Phase diagram of helically imbalanced QCD matter}

\author{M. N. Chernodub}
\affiliation{Institut Denis Poisson, Universit\'e de Tours, Tours 37200, France}
\affiliation{Pacific Quantum Center, Far Eastern Federal University, Sukhanova 8, Vladivostok, 690950, Russia}
\author{Victor E. Ambru\cb{s}}
\affiliation{
Department of Physics, West University of Timi\cb{s}oara,
Bd.~Vasile P\^arvan 4, Timi\cb{s}oara 300223, Romania}

\date{\today}

\begin{abstract}
We discuss the influence of a helicity imbalance on the phase diagram of dense QCD at finite temperature. We argue that the helical chemical potential is a thermodynamically relevant quantity in theories with the mass gap generation. Using the linear sigma model coupled to quarks, we show that the presence of the helical density substantially affects the phase diagram of dense quark matter. A moderate helical density makes the chiral phase transition softer while shifting the critical endpoint towards lower temperatures and higher baryon chemical potentials. As the helical density increases, the segment of the first-order transition disappears, and the chiral transition becomes a soft crossover. At even higher helical chemical potentials, the first-order transition reappears again at the zero-density finite-temperature transition and extends into the interior of the phase diagram. This evolution of the chiral transition reflects the existence of a thermodynamic duality between helical and vector (baryonic) chemical potentials. We also show that the presence of the helicity imbalance of quark matter increases the curvature of the chiral pseudocritical line in QCD.
\end{abstract}

\pacs{12.38.Aw, 25.75.Nq, 12.38.Mh}

\maketitle

\section{Introduction}

Unusual properties of quark-gluon plasma attract intensive attention of the scientific community. Nowadays, this ultrahot state of matter is routinely created in relativistic heavy-ion collisions~\cite{Gyulassy:2005npa,Muller:2012ann,Jacak:2012nature} thus making it possible to probe experimentally its thermodynamics, phase diagram, equation of state, as well as various transport phenomena~\cite{Romatschke:2007prl,Heinz:2013ar,Ryu:2015prl}. Recently, the ultra-peripheral collisions opened the door to the investigation of the highly-rotating plasma seen experimentally via the quarks' spin degrees of freedom~\cite{STAR:2017ckg,Becattini:2020ngo}.

The spin degree of freedom of an ultrarelativistic quark can be quantified via its helicity $h = {\bs s} \cdot {\bs p}/|\p|$, which is the projection of the quark's spin $\bs s$ onto the quark's momentum $\bs p$. The definition of helicity $h$ applies in exactly the same way both to quarks and to anti-quarks. One distinguishes the right- and left-handed quarks with, respectively, positive and negative values of the helicity~\cite{Pal:2010ih}.

The notion of helicity is usually used as an intermediate step to describe the physical sense of a very similar quantity, called chirality. For a Dirac fermion, the chirality is even under the charge conjugation ($C$) transformation, while the helicity is odd. For example, a quark with a right-handed helicity has a right-handed chirality while an anti-quark with the very same right-handed helicity has an opposite, left-handed chirality. The chirality is determined as an eigenvalue of the fifth gamma matrix~$\gamma^5$. 

In the context of QCD, the transformations generated by the matrix $\gamma^5$ are usually associated with the ``axial'' $U(1)_A$ subgroup of a larger group of global QCD symmetries (the latter group carries the very name ``chiral''). Therefore below we will use mostly the term ``axial symmetry'' simultaneously referring to the ``chirality'' of quarks.

The importance of the axial symmetry is determined by its significant influence on the properties of QCD, in particular, to the topological structure of the QCD vacuum. The axial symmetry, which is respected by the massless Dirac Hamiltonian, is broken at the quantum level via an axial anomaly. This feature leaves an imprint on the particularities of the meson spectrum~\cite{ref:meson} and generates anomalous transport effects in the quark-gluon plasma (QGP) created in relativistic heavy-ion collisions~\cite{Landsteiner:2011prl,Kharzeev:2013ffa}. The axial density of quarks modifies the thermodynamic properties of the plasma and its phase diagram~\cite{Ruggieri:2011xc,Chernodub:2011fr,Gatto:2011wc,Ruggieri:2016ejz,Frasca:2016rsi,Astrakhantsev:2019wnp}.

While the axial properties of QCD are discussed in great details, the helical quantum numbers have not been studied with a due attention. Despite the chirality and helicity being very similar to each other, they, nevertheless, possess quite different features. For example, at a classical level, the axial charge is conserved only for massless fermions, while the helical charge is conserved for any value of the fermion mass. The axial charge is determined with the help of a local Lorentz-invariant operator, while the definition of the helical charge relies on the local frame (the latter feature, however, is not important for theories at finite density and/or temperature). 

One could also expect the existence of similarities between the axial and helical quantum numbers. On the quantum level, the helical degrees of freedom -- similarly to their axial counterparts~\cite{Kharzeev:2013ffa,Kharzeev:2016ppnp} -- may lead to new nondissipative transport phenomena, the helical vortical effects, that emerge in a helically-imbalanced rotating fermionic system~\cite{Ambrus:2019helican}. Both chirality~\cite{Ruggieri:2016asg} and helicity~\cite{Kapusta:2019ktm} may have the equilibration times close to the relaxation time of the spin degrees of freedom~\cite{Kapusta:2020npk}. 

In our paper, we discuss the influence of a global helical charge density on thermodynamics of strong interactions. We assume that the helical charge, similarly to the axial charge, may be generated due to thermal fluctuations of a non-equilibrium environment at the early stages of heavy ion collisions. To address the thermalized phase, we use the effective approach based on the linear sigma model coupled to quarks (\M) which also serves as an effective low-energy model of QCD~\cite{ref:LSMq}. 

As we discuss at the end of the paper, the net helicity is expected to be a reasonably good quantum number to characterize the thermal evolution of the quark-gluon plasma until the hadronization time. Indeed, it is well-known that the helicity of massless quarks is conserved in perturbative QCD interactions due to the vector nature of the coupling between quarks and gluons (see, for example, the discussions in Refs.~\cite{Kapusta:2019sad,Kapusta:2019ktm,Kapusta:2020npk}). This statement is applied, in particular, to the high-temperature phase of QCD where the light quark masses are small compared to their thermal energy.

The structure of our paper is as follows. In Section~\ref{sec:helicity}, we discuss differences and similarities between the thermodynamics of vector, axial, and helical charges and corresponding chemical potentials. Surprisingly, we find that the helical density is closer to the vector density rather than to its axial counterpart. We recall, after Ref.~\cite{ref:Marco}, how the presence of mass for free fermions makes the axial chemical potential thermodynamically inconsistent. We demonstrate that the helical chemical potential does not possess this property. We describe the linear sigma model coupled to quarks (\M) in Section~\ref{sec:model}. We use this model to discuss, in Section~\ref{sec:phase}, the thermodynamics of the dense QCD matter in the presence of the helical chemical potential. We calculate the phase diagram of the model and study the evolution of the chiral transition as the helical chemical potential increases. The last section is devoted to our conclusions.

\section{Chirality and helicity in thermodynamics of free fermions}
\label{sec:helicity}

Before going into the details of QCD thermodynamics, let us discuss first the role of chirality and helicity in the thermodynamic properties of free Dirac fermions.

\subsection{Chirality vs. Helicity for Dirac fermions}

Consider free massive Dirac fermions with the Lagrangian:
\beqn
\cL = \overline{\psi} \left(i \slashed\partial + m\right)\psi\,,
\label{eq:L:free}
\eeqn
where we use the slashed notation $\slashed\partial = \gamma^\mu \partial_\mu$ expressed via the Dirac $\gamma^\mu$ matrices ($\mu = 0, \dots, 3$), and ${\bar \psi} = \psi^\dagger \gamma^0$. We will also use the fifth gamma matrix, $\gamma^5 = i \gamma^0 \gamma^1 \gamma^2 \gamma^3$.

The axial charge (chirality) $\chi = \pm 1$ of a fermion state $\psi$ is defined according to an eigenvalue $\chi$ of the $\gamma^5$ matrix, $\gamma^5 \psi = \chi \psi$. One distinguishes the right-handed ($R$) and left-handed ($L$) chiral eigenstates, respectively:
\beqn
\gamma^5 \psi_R = + \psi_R, 
\qquad
\gamma^5 \psi_L = - \psi_L.
\label{eq:chirality}
\eeqn

As we mentioned in the Introduction, the chirality $\chi$ of a fermion state is closely related to the helicity $\lambda$ of the same state. Classically, the helicity is determined by the projection of the spin $\bs s$ on the direction of motion of the fermion given by its momentum $\bs p$. At the quantum level, the helicity $\lambda$ is an eigenvalue of the helicity operator:
\beqn
h = \frac{{\bs s}\cdot {\bs p}}{p} 
\equiv
\frac{\gamma^5 \gamma^0}{2} \frac{{\bs \gamma} \cdot \p}{|\p|}, 
\label{eq:h}
\eeqn
where ${\bs p} = - i {\bs \partial}$ is the momentum operator, $p = |{\bs p}|$ is its absolute value, and $s_i = \frac{1}{2} \varepsilon_{0ijk} \Sigma^{jk}$ is the spin operator which is constructed from the covariant antisymmetric tensor $\Sigma^{\mu\nu} = \frac{i}{4}[\gamma^\mu,\gamma^\nu]$. 

Since the fermion is a spin $1/2$ particle, the helicity operator~\eq{eq:h} takes two values, $\pm 1/2$. It is convenient to rescale, by the factor of two, both the helicity operator $\h = 2 h$ and the corresponding helicity eigenvalue $\varkappa$ with $\h \psi = \varkappa \psi$. The rescaled helicity operator $\h$ has the convenient eigenvalues $\pm 1$. One distinguishes the right-handed ($\uparrow$) and the left-handed ($\downarrow$) helicity eigenstates:
\begin{equation}
 \h\psi_{\uparrow} = + \psi_{\uparrow}, \qquad
 \h\psi_{\downarrow} = -\psi_{\downarrow}.
\label{eq:helicity}
\end{equation}

At the level of the classical Dirac equation, it can be easily seen that the helicity is a conserved quantity, as follows. Consider the Dirac equation, $(i \slashed{\partial} - m) \psi$, in the following form:
\begin{equation}
    i \partial_t \psi = H \psi,\qquad 
    H = -i \gamma^0 \bm{\gamma} \cdot \bm{\nabla}  + m \gamma^0,
\end{equation}
where $H$ is the Hamiltonian of the system.
In order to be conserved, the helicity should satisfy
\begin{equation}
 \h i \partial_t \psi = i \partial_t (\h \psi),
\end{equation}
or equivalently, $[\h, H] = 0$. This latter equality is readily checked by noting that:
\begin{align}
    \h H =& \gamma^5 \gamma^0 \left(-i \
    \frac{\bm{\gamma} \cdot \bm{\nabla}}{|\bm{p}|}\right) \left(-i \gamma^0 \bm{\gamma} \cdot \bm{\nabla}  + m \gamma^0\right)
    \nonumber\\
    =& \gamma^5 \gamma^0 \left(i \gamma^0 \bm{\gamma} \cdot \bm{\nabla}  - m \gamma^0\right)
    \left(-i \
    \frac{\bm{\gamma} \cdot \bm{\nabla}}{|\bm{p}|}\right) \nonumber\\
    =& H \h.
    \label{eq:h:H}
\end{align}

The chirality and helicity are different quantities. For a single particle, these quantum numbers are firmly related to each other: The chirality of a particle is equal to its helicity (for example, a right-chiral particle has a right-handed helicity) while the chirality of an antiparticle is opposite to its helicity (for instance, a right-chiral antiparticle has a left-handed helicity). However, the total helicity of an ensemble of particles cannot be determined only by its total vector charge and total axial charge. Therefore the helicity, given its conservation for free massive fermions, may serve -- in addition to a vector (baryonic) charge -- as a useful quantity to characterize the thermodynamic ensembles of fermions. 

We would like to stress that
it is important for us to consider the theory with massive fermions in view of its further applicability to QCD. Although the mass gap generation emerges at the purely gluon sector of QCD, this phenomenon is accompanied by the chiral symmetry breaking at the quark sector which gives masses to quarks via a dynamical mechanism \cite{Fukushima:2011rpp}. In the next section, we discuss thermodynamics of free massive fermions for a number of chemical potentials. First, we consider the well-known case of the vector (related to baryonic) chemical potential. Then we show, following Ref.~\cite{ref:Marco}, that the presence of non-zero fermionic mass is absolutely inconsistent with the presence of a finite axial chemical potential starting at the level of zero-point fluctuations. Finally, we discuss the helical chemical potential and show its consistency with the mass gap generation.

\subsection{Thermodynamics of free fermions with vector, axial, and helical chemical potentials}

\subsubsection{General formalism}

A free fermion with the mass $m$ in the presence of the vector ($\mu_V$), axial ($\mu_A$), and helical ($\mu_H$) chemical potentials can be described by the following effective Lagrangian \cite{Laine:2016}:
\beqn
\cL = \overline{\psi} \left(i \slashed\partial + \mu_V \gamma^{0} + \mu_A \gamma^{0}\gamma^{5} + \mu_H \gamma^{0} \h - m\right)\psi\,.
\label{eq:L:free:All:mu}
\eeqn

It is convenient to rewrite the corresponding Dirac equation,
\beqn
\left(i \slashed\partial + \mu_V \gamma^{0} + \mu_A \gamma^{0}\gamma^{5} + \mu_H \gamma^{0} \h - m\right) \psi = 0,
\label{eq:Dirac}
\eeqn
in terms of the plane waves: 
\beqn
\psi(x) = \chi_p\, e^{- i p_\mu x^\mu},
\label{eq:plane:wave}
\eeqn
where $x = (t, \x)$ and $p^\mu = (p_0, \p)$ and the momentum-dependent spinor $\chi_p$. 
We use the flat metric with the $(+,-,-,-)$ signature. In the momentum space, the Dirac equation~\eq{eq:Dirac} reduces to the set of linear equations:
\beqn
\cM(p)\chi_p = 0,
\label{eq:Dirac:p}
\eeqn
determined by the following matrix:
\beqn
\cM(p) = \slashed p + \mu_V \gamma^{0} + \mu_A \gamma^{0}\gamma^{5} + \mu_H \gamma^{0} \h - m.
\label{eq:Mp}
\eeqn
A consistent solution of Eq.~\eq{eq:Dirac} requires the determinant of the matrix~\eq{eq:Mp} to vanish. This condition leads to a polynomial equation:
\beqn
\det \cM(p) = 0,
\label{eq:cM:0}
\eeqn
which has four roots in terms of the zeroth component of the momentum~$p_0= p^{(s)}_{0,\varkappa}(\p)$:
\beqn
\det \cM(p) = \prod_{\varkappa = \pm 1} \prod_{s = \pm}
\left[ p_0 - p^{(s)}_{0,\varkappa}(\p) \right].
\label{eq:cM:prod}
\eeqn
The roots $p_0$ are labeled by the helicity $\varkappa = \pm 1$, determined via Eq.~\eq{eq:helicity}, and the kind $s=\pm 1$ of the solution for particle $(s=+)$ and anti-particle $(s=-)$ energy branches. The solutions of Eq.~\eq{eq:cM:prod} depend on the spatial momentum $\p$, the mass $m$, and the full set of chemical potentials,
$\left( \mu_V, \mu_A, \mu_H \right)$, being given by
\begin{equation}
 p^{(s)}_{0,\varkappa}(\p) = -\mu_V - \varkappa \mu_H 
 + s \sqrt{m^2 + (|\p| - \varkappa \mu_A)^2}.
\end{equation}

Although we call the variety of these solutions as the ``energy branches'', the quantity $p^{(s)}_{0,\varkappa}(\p)$ does not have the literal sense of energy. For example, the condition $p^{(s)}_{0,\varkappa}(\p) = 0$ defines, depending on the existence of the real-valued solution, the position of the Fermi surface for particle ($s=+1$) or anti-particle ($s=-1$) states of fermions carrying the helicity $\varkappa$. 

It is convenient to compute the free energy of the Dirac system in the Euclidean spacetime after performing the Wick rotation, $p_0 \to i p_4 \to i \varpi_n$:
\beqn
\Omega = - T \sum_{n\in \Z}\int \frac{d^3 p}{(2\pi)^3} \ln \det \frac{\cM (p)}{T} {\biggl|}_{p_0 = i \varpi_n}
\label{eq:Omega:1}
\eeqn
where $\varpi_n = \pi T (2n + 1)$ is the fermionic Matsubara frequency at temperature $T$ labeled by the index $n \in \Z$ \cite{Laine:2016}.

The free energy~\eq{eq:Omega:1} may be rewritten using Eq.~\eq{eq:cM:prod}:
\beqn
\Omega = - T \sum_{\varkappa = \pm 1} \sum_{s = \pm} \sum_{n\in \Z}\int \frac{d^3 p}{(2\pi)^3} \ln  \frac{\varpi_n + i s p^{(s)}_{0,\varkappa}(\p)}{T},\qquad
\label{eq:Omega:2}
\eeqn
where the additional multiplier $s$ takes into account the correct contour of integration along the momentum $p_0$. After the Wick rotation, the integration becomes a sum over the Matsubara frequencies $\varpi_n$ in the Euclidean representation of the free energy~\eq{eq:Omega:2}.

We take into account the identity
\beqn
\ln  \frac{\varpi_n + i p_0}{T} = i \int_0^{p_0/T}
\frac{d \theta}{\pi(2n+1) + i \theta} + C_n,
\label{eq:int:1}
\eeqn
and neglect the inessential constant $C_n = \ln \pi (2n+1)$ in the following. The summation over $n$ in Eq.~\eq{eq:Omega:2} can be performed with the help of Eq.~\eq{eq:int:1} and the following relation:
\beqn
\sum_{n \in \Z} \frac{1}{\varpi_n + i p_0} = \frac{i}{T} 
\left[ n_T(p_0) - \frac{1}{2}\right]
\eeqn
where 
\beqn
n_T(\omega) = \frac{1}{e^{\omega/T} + 1}
\eeqn
is the Fermi-Dirac distribution. The integral over the variable $\theta$ may be taken using the identity:
\beqn
\int_0^\x \frac{d \theta}{e^\theta + 1} =
- \ln \left(1 + e^{-x} \right) + \ln 2\,.
\eeqn
Below we will again neglect an inessential constant $\ln 2$.

Finally, we get the following expression for the free energy:
\beqs
\beqn
\Omega\ & = & \Omega_\ZP + \Omega_T, \\
\Omega_\ZP & = & - \frac{1}{2} \sum_{\varkappa = \pm 1} \sum_{s = \pm} \int \frac{d^3 p}{(2\pi)^3} s p^{(s)}_{0,\varkappa}(\p),
\label{eq:Omega:0}\\
\Omega_T & = & - \sum_{\varkappa = \pm 1} \sum_{s = \pm} \int \frac{d^3 p}{(2\pi)^3} T \ln \left(1 + e^{- s p^{(s)}_{0,\varkappa}(\p)/T} \right).\qquad
\label{eq:Omega:T}
\eeqn
\label{eq:Omega:3}
\eeqs
This expression is the most convenient representation of the free energy $\Omega_T$ as it contains all four branches of the energy dispersion. The zero-point term $\Omega_{\mathrm{ZP}}$ is usually associated with the vacuum contribution while the term $\Omega_T$ represents the thermal and matter contributions to the free energy.

The densities of all three charges which can be obtained via the differentiation of the free energy~\eq{eq:Omega:3} with respect to the corresponding chemical potential:
\beqn
n_\ell = - \frac{1}{\mathrm{Vol}} \frac{\partial \Omega}{\partial \mu_\ell}, 
\qquad
\ell = V, A, H.
\label{eq:Q:ell}
\eeqn
These densities correspond to the vacuum expectation values of the zero components, $n_\ell = \avr{J^0_\ell}$, of the vector, axial, and helical currents, respectively,
\beqn
J^\mu_V =  \overline{\psi} \gamma^\mu \psi, \quad 
J^\mu_A =  \overline{\psi} \gamma^\mu \gamma^5 \psi, \quad
J^\mu_H =  \overline{\psi} \gamma^\mu \h \psi. \qquad \label{eq:J}
\eeqn
These currents form a ``triad'' of classically conserved $U(1)$ quantities for massless ($m=0$) Dirac fermions. In this article, however, we will be interested in properties of quarks with a dynamically generated mass. One can check that the vector and helical charges are still classically conserved as the classical solutions of massive fermions~\eq{eq:L:free:All:mu} satisfy the equations $\partial_\mu J^\mu_V = \partial_\mu J^\mu_H = 0$ identically. We will see that the fact that the axial charge is not conserved for massive fermions, $\partial_\mu J^\mu_A \neq 0$, will profoundly affect the thermodynamics of fermions with the axial chemical potential.

Below we discuss the effects of each chemical potential on the thermodynamics of the system. In order to get a clear picture, we consider a single nonzero chemical potential and require that the other two vanish. 

\subsubsection{Vector chemical potential ($\mu_V \neq 0$, $\mu_A = \mu_H = 0$)}

First we consider the well-known case with a finite vector density. The Dirac Lagrangian with the vector chemical potential $\mu_V$,
\beqn
\cL_V = {\bar \psi}\left(i \slashed\partial + \mu_V \gamma^{0} - m\right) \psi,
\eeqn
describes particles, for which the temporal $p_0$ and spatial components $\bs p$ of the four-momentum are related, via  Eq.~\eq{eq:cM:0}, as follows:
\beqn
p^{(s)}_{0,\varkappa}(\p) = -\mu_V + s \sqrt{\p^2 + m^2}\,.
\label{eq:energy:muV}
\eeqn
The vector chemical potential $\mu_V$ shifts the particle $(s=+1)$ and anti-particle $(s=-1)$ energy branches by the same value of energy $\mu_V$ which does not depend neither on particle type $s$ nor on the helicity quantum number $\varkappa$. We find that each of the levels~\eq{eq:energy:muV} is double-degenerate with respect to helicity $\varkappa$.

Using the dispersion relations~\eq{eq:energy:muV}, the free energy~\eq{eq:Omega:3} can be represented as the sum
\beqn
\Omega^V(T,\mu_V) = \Omega^V_{\mathrm{vac}} + \Omega^V_T(T,\mu_V)\,.
\eeqn
The total free energy contains the divergent vacuum part
\beqn
\Omega^V_{\mathrm{vac}} \equiv \Omega^V_\ZP = - 2 \int \frac{d^3 p}{(2\pi)^3} \omega_\p(m),
\label{eq:Omega:0:V}
\eeqn
and the finite thermodynamic contribution:
\beqn
\Omega^V_T = - 2 T \sum_{s = \pm 1} \int \frac{d^3 p}{(2\pi)^3}  \ln \left(1 + e^{- \frac{\omega_\p - s \mu_V}{T}} \right),\quad
\label{eq:Omega:T:V}
\eeqn
where 
\beqn
\omega_\p = \sqrt{\p^2 + m^2},
\label{eq:omega:p}
\eeqn
is the one-particle energy. The vacuum part~\eq{eq:Omega:0:V} does not contribute to the thermodynamics of the system as it depends neither on temperature $T$ nor on the chemical potential $\mu_V$.

The density of the vector (``electric'') charge is then given by the thermodynamic part~\eq{eq:Omega:T:V} with the help of Eq.~\eq{eq:Q:ell} with $\ell = V$:
\beqn
n_V = 2 \int \frac{d^3 p}{(2\pi)^3} \left( \frac{1}{e^{ \frac{\omega_\p - \mu_V}{T}} +1} - \frac{1}{e^{ \frac{\omega_\p + \mu_V}{T}}+1} \right).
\label{eq:nV:T}
\eeqn
The vacuum part~\eq{eq:Omega:0:V} does not contribute to the density.

At small mass, the explicit integration in Eq.~\eq{eq:nV:T} gives \cite{Ambrus:2019jhep}:
\beqn
n_V(T,\mu_V) = \frac{\mu_V T^2}{3} + 
\frac{\mu_V^3}{3\pi^2} - \frac{\mu_V m^2}{2\pi^2} + O(m^4).
\label{eq:QV:m0}
\eeqn

\subsubsection{Axial chemical potential ($\mu_A \neq 0$, $\mu_V = \mu_H = 0$)}

The Dirac particles at the axial chemical potential $\mu_A$ is described by the Lagrangian:
\beqn
\cL_A = {\bar \psi}\left(i \slashed\partial + \mu_A \gamma^{0} \gamma^5 - m\right) \psi.
\eeqn
For simplicity of our analysis, we keep the vector and helical chemical potentials vanishing, $\mu_V = \mu_H = 0$.

Repeating all the steps of the previous section, we find that in the present case, the energy dispersions, constrained by the relation~\eq{eq:cM:0}, are as follows:
\beqn
p^{(s)}_{0,\varkappa}(\p) = s \sqrt{(|\p| - \varkappa \mu_A)^2 + m^2}.
\label{eq:energy:muA}
\eeqn
These states are characterized by the particle/antiparticle number $s = \pm 1$ and the helicity $\varkappa = \pm 1$.

In a sharp contrast with Dirac fermions at a nonzero vector charge density, the would-be vacuum term $\Omega_\ZP^A$~\eqref{eq:Omega:0} of the $\mu_A \neq 0$ fermions depends explicitly on the axial chemical potential $\mu_A$. Indeed, when $p^{(s)}_{0,\varkappa}(\p)$ is given by Eq.~\eqref{eq:energy:muA}, $\Omega^A_\ZP$ contains the truly vacuum part $\Omega^A_{\mathrm{vac}}$, which is equal to $\Omega^V_{\mathrm{vac}}$ given in Eq.~\eqref{eq:Omega:0:V}, as well as a ``density'' part that depends on the axial chemical potential, $\Omega^A_{\mathrm{dens}}$:
\beqn
\Omega^A_\ZP = \Omega^A_{\mathrm{vac}} + \Omega^A_{\mathrm{dens}}.
\label{eq:Omega:A:ZP}
\eeqn
Thus we divide the free energy~\eq{eq:Omega:3} into the following three terms:
\beqn
\Omega^A(T,\mu_A) = \Omega^A_{\mathrm{vac}} + \Omega^A_{\mathrm{dens}}(\mu_A) + \Omega^A_T(T,\mu_A)\,.
\label{eq:Omega:A:tot}
\eeqn

The finite-density part in Eq.~\eq{eq:Omega:A:tot}:
\beqn
\Omega^A_{\mathrm{dens}}(\mu_A) & = & - \sum_{\varkappa = \pm 1} \int \frac{d^3 p}{(2\pi)^3} \\
& & \cdot \left(\sqrt{(|\p| - \varkappa \mu_A)^2 + m^2} - \sqrt{\p^2 + m^2}\right),  \nonumber
\label{eq:Omega:dens:A}
\eeqn
also comes from the ``vacuum fluctuation'' term $\Omega_\ZP$, which gets this unconventional (and, as we show below, somewhat artificial) contribution.

The thermal contribution to the free energy is given by the following finite expression:
\beqn
\Omega^A_T = - 2 T \!\!\sum_{\varkappa = \pm 1} \int \frac{d^3 p}{(2\pi)^3}  \ln \left(1 + e^{- \frac{\sqrt{(|\p| - \varkappa \mu_A)^2 + m^2}}{T}} \right).\qquad
\label{eq:Omega:T:A}
\eeqn
Its form is somewhat unusual due to the fact that the dispersion relation in Eq.~\eq{eq:Omega:T:A} is different from the conventional one-particle dispersion relation~\eq{eq:omega:p}.

The appearance of the finite-density part~\eq{eq:Omega:dens:A} has the self-contradictory ``vacuum'' origin. This term determines the axial density~\eq{eq:Q:ell} of Dirac fermions at vanishing temperature: 
\beqn
& & n_A(\mu_A) \biggl{|}_{T = 0} = - \frac{\partial \Omega^A_{\mathrm{dens}}(\mu_A)}{\partial \mu_A}
\nonumber \\
& & \quad = \sum_{\varkappa = \pm 1} \int \frac{d^3 p}{(2\pi)^3} \frac{\mu_A - \varkappa |\p|}{\sqrt{(\mu_A - \varkappa |\p|)^2 + m^2}}, 
\label{eq:Q:A:dens}
\eeqn

The axial density~\eq{eq:Q:A:dens} for massless fermions ($m=0$) has a conventional, non-divergent expression:
\beqn
n_A(\mu_A) \biggl{|}_{{}^{T = 0}_{m = 0}} =
2 \int \frac{d^3 p}{(2\pi)^3} \Theta(\mu_A - p)
= \frac{\mu_A^3}{3 \pi^2},
\label{eq:nA:T0}
\eeqn
where $\Theta(x)$ is the Heaviside step function.

It's remarkable to notice that the cutoff in Eq.~\eq{eq:nA:T0} at the Fermi-momentum $p = \mu_A$ appears not in the thermodynamic part~\eq{eq:Omega:T:A} -- which is always zero for $T=0$ and $\mu_A \neq 0$ -- but it comes naturally in the ``vacuum'' contribution. 

At finite temperature, the contribution $n_{A;T} = -\partial \Omega^A_T / \partial \mu_A$ from $\Omega^A_T$ to the axial charge density is 
\begin{multline}
 n_{A;T} = \frac{1}{\pi^2} 
 \sum_{\varkappa = \pm 1} \int_0^\infty dp \frac{ p^2(\varkappa p - \mu_A)}{\sqrt{(\varkappa p - \mu_A)^2 + m^2}}\\
 \times \left\{ 
 \exp\left[\frac{1}{T} \sqrt{(\varkappa p - \mu_A)^2 + m^2}\right] + 1\right\}^{-1}.
\end{multline}
Adding now the vanishing temperature contribution $n_A(\mu_A)\biggl|_{T = 0}$, coming from $\Omega_{\rm dens}^A$, we obtain
\begin{multline}
 n_A(T) = -\frac{1}{2\pi^2} 
 \sum_{\varkappa = \pm 1} \int_0^\infty dp \frac{ p^2(\varkappa p - \mu_A)}{\sqrt{(\varkappa p - \mu_A)^2 + m^2}}\\
 \times \tanh\left[\frac{1}{2T} \sqrt{(\varkappa p - \mu_A)^2 + m^2}\right].
\end{multline}
At vanishing mass, the above expression simplifies to
\begin{align}
 n_A(T)\biggl|_{m = 0} =& -\frac{1}{2\pi^2} 
 \sum_{\varkappa = \pm 1} \int_0^\infty dp \, p^2 \,
 \tanh\frac{\varkappa p - \mu_A}{2T}
 \nonumber\\
 =& \frac{1}{\pi^2} 
 \sum_{\varkappa = \pm 1} \int_0^\infty dp \, p^2 
 \frac{\varkappa}{e^{(p - \varkappa \mu_A) / T} + 1} \nonumber\\
 =& \frac{\mu_A T^2}{3} + \frac{\mu_A^3}{3\pi^2}.
\end{align}

For massive fermions, however, the interpretation of the axial density, generated by the unexpected ``vacuum'' contribution~\eq{eq:Q:A:dens}, becomes less clear~\cite{ref:Marco}.
For example, consider the axial density at high chemical potential ($\mu_A \gg m$) for massive fermions. At high momenta, $|\p| \gg \mu_A$, the expression under the integral~\eq{eq:Q:A:dens} vanishes as fast as $2 m^2 \mu_A/|\p|^3$ which is not, however, enough to make the whole integral convergent. In fact, the axial density diverges logarithmically in the ultraviolet region:
\beqn
n_A(\mu_A) \biggl{|}_{{}^{T = 0}_{m \ll |\mu_A|}} = \frac{\mu_A^3}{3 \pi^2} + \frac{m^2 \mu_A}{\pi^2} \ln \frac{\Lambda_{\mathrm{UV}}}{m} + \dots,
\label{eq:nA:div}
\eeqn
where the ellipsis indicate non-divergent terms of the order of $O(\mu_A)$ and $\Lambda_{\mathrm{UV}}$ indicates the ultraviolet cutoff. 

The logarithmic divergence of the axial density~\eq{eq:nA:div} appears as a result of the lack of axial symmetry for massive Dirac fermions~\cite{ref:Marco}. The axial density $Q_A$ is not a conserved quantity if the Dirac fermions are massive. Indeed, the chemical potential cannot be introduced self-consistently for a non-conserved charge. Therefore, the presence of both $\mu_A \neq 0$ and $m \neq 0$ cannot be set in a physically self-consistent manner.  

The physical situation becomes even more subtle in the case of theories where the mass is generated dynamically, as it happens, for example, in interacting field theories such as QCD. In this case, the axial chemical potential may lead to an additional renormalization which is discussed in details in Ref.~\cite{ref:Marco}. Basically, the infinite zero-point energy cannot be removed  by the usual subtraction procedure as it contains both the vacuum part and the contribution coming from matter~\eq{eq:Omega:A:ZP}.

\subsubsection{Helical chemical potential ($\mu_H \neq 0$, $\mu_V = \mu_A = 0$)}

Finally, we consider the helical chemical potential which is the central topic of our paper. In the theory of free massive fermions, the helical potential $\mu_H$ appears in the Lagrangian as follows:
\beqn
\cL = \overline{\psi} \left(i \slashed\partial + \mu_H \gamma^0 \h - m\right)\psi\,.
\label{eq:L:free:muH}
\eeqn
We keep the vector and axial chemical potentials equal to zero, $\mu_V = \mu_A =0$.

The energy dispersion condition~\eq{eq:cM:0} for the Lagrangian~\eq{eq:L:free:muH} 
give us the following four energy branches:
\beqn
p^{(s)}_{0,\varkappa}(\p) = s \sqrt{\p^2 + m^2} - \varkappa \mu_H\,.
\label{eq:energy:muH}
\eeqn
Thus, the helical chemical potential $\mu_H$ shifts the particle $(s=+1)$ and anti-particle $(s=-1)$ branches with the energy which has the standard form~\eq{eq:omega:p}. The sign of the shift now depends explicitly on the helicity $\varkappa$ of the branch, as one could expect from a quantity which is invoked to distinguish the helicity.

Surprisingly, the effects of nonzero vector~\eq{eq:energy:muV}  and helical~\eq{eq:energy:muH} chemical potentials on the energy branches are quite similar to each other in the sense that both potentials shift the spectra without modifying the functional dependence of the energy on momentum. Both vector and helical potentials differ significantly from the axial chemical potential, which alters the very form of the energy levels~\eq{eq:energy:muA}. 

The free energy in the presence of the helical chemical potential contains two terms:
\beqn
\Omega^H(T,\mu_A) = \Omega^H_\ZP + \Omega^H_T(T,\mu_A)\,.
\label{eq:Omega:H:tot}
\eeqn
The zero-point fluctuations lead to the conventional vacuum term~\eq{eq:Omega:0:V} $\Omega^H_\ZP = \Omega^H_{\mathrm{vac}} = \Omega^V_{\rm vac}$, which is independent of temperature and chemical potential.
The thermodynamic part of the free energy is
\beqn
\Omega^H_T(\mu_H) & = & - T \sum_{s = \pm 1} \sum_{\varkappa = \pm 1}\int \frac{d^3 p}{(2\pi)^3}  \nonumber \\
& & \hskip 12mm \cdot \ln \left(1 + e^{- \frac{\omega_\p - s \varkappa \mu_H}{T}} \right).\quad\
\label{eq:Omega:T:H}
\eeqn

Remarkably, the dependence of the free energy~\eq{eq:Omega:T:H} on the helical chemical potential $\mu_H$ mimics exactly the one~\eq{eq:Omega:T:V} of the vector chemical potential $\mu_V$:
\beqn
\Omega_H(\mu_H) = \Omega_V(\mu_V) {\biggl|}_{\mu_V \to \mu_H}. 
\label{eq:Omega:H:V:equiv}
\eeqn
Of course, the relation~\eq{eq:Omega:H:V:equiv} does not mean that the effects of the helical and vector potentials on the Dirac fermions are identical to each other: it is the parametric dependence of the free energy which is the same in both cases. In order to demonstrate this fact, we will consider, in the next subsection, the free energy of Dirac fermions in the presence of both these chemical potentials. Meanwhile, we give the explicit expression for the helical density \cite{Ambrus:2019jhep}:
\beqn
n_H(T,\mu_H) = \frac{\mu_H T^2}{3} + \frac{\mu_H^3}{3\pi^2} - \frac{\mu_H m^2}{2\pi^2} + O(m^4).
\label{eq:nH:m0}
\eeqn

\subsubsection{Duality of helical and vector chemical potentials}

Now we consider the thermodynamics of Dirac fermions in the presence of both vector and helical densities. This physical environment is described by the Lagrangian:
\beqn
\cL = \overline{\psi} \left(i \slashed\partial + \mu_V \gamma^{0} + \mu_H \gamma^{0} \h - m\right)\psi\,,
\label{eq:L:free:VH}
\eeqn
which gives the following Dirac equation:
\beqn
\left(i \slashed\partial + \mu_V \gamma^{0} + \mu_H \gamma^{0} \h - m\right) \psi = 0.
\label{eq:Dirac:VH}
\eeqn

The spectrum is described by the four energy branches:
\beqn
p^{(s)}_{0,\varkappa}(\p) = s \sqrt{\p^2 + m^2} - \mu_V - \varkappa \mu_H\,,
\label{eq:energy:muVH}
\eeqn
which are immediate generalizations of the vector~\eq{eq:energy:muV} and helical~\eq{eq:energy:muH} energy solutions.

The vacuum contribution to the free energy may traditionally be neglected below as it depends neither on temperature nor on chemical potentials. The thermodynamic contribution is as follows:
\beqn
\Omega^{VH}_T(\mu_V,\mu_H) & = & - T \sum_{s = \pm 1} \sum_{\varkappa = \pm 1}\int \frac{d^3 p}{(2\pi)^3}  \nonumber \\
& & \times \ln \left(1 + e^{- \frac{\omega_\p - 
s (\mu_V + \varkappa \mu_H)}{T}} \right),\quad\
\label{eq:Omega:T:VH}
\eeqn
The form of the thermodynamic potential~\eq{eq:Omega:T:VH} demonstrates the independence of the physical effects of vector and helical chemical potentials. The potentials appear to enter the partition function symmetrically, exhibiting the symmetry of thermodynamic function $\Omega_T \equiv \Omega^{VH}_T$ under the flip of the chemical potentials \beqn
\left(
\begin{array}{c}
     \mu_V  \\
     \mu_H
\end{array}
\right)
\to 
\left(
\begin{array}{c}
     \mu_H  \\
     \mu_V
\end{array} 
\right)
\label{eq:duality}
\eeqn
namely: 
\beqn
\Omega_T(\mu_V,\mu_H)
= \Omega_T(\mu_H,\mu_V).
\label{eq:VH:symmetry}
\eeqn
The free energy~\eq{eq:Omega:T:VH} depends on the absolute values and not on the signs of the chemical potentials $\mu_V$ and $\mu_H$. Thus, the thermodynamics of the theory is also invariant under the sign flips $\mu_V \to \pm \mu_V$ and $\mu_H \to \pm \mu_H$. In the small mass limit, $n_V$ and $n_H$ are given by:
\begin{align}
 n_V =& \frac{\mu_V T^2}{3} + 
\frac{\mu_V^3 + 3 \mu_V \mu_H^2}{3\pi^2} - \frac{\mu_V m^2}{2\pi^2} + O(m^4),\nonumber\\
 n_H =& \frac{\mu_H T^2}{3} + 
\frac{\mu_H^3 + 3 \mu_H \mu_V^2}{3\pi^2} - \frac{\mu_H m^2}{2\pi^2} + O(m^4).
\end{align}

We conclude this section by stressing that the presence of vector and helical chemical potentials and the appropriate densities is, expectedly, consistent with the thermodynamics of the massive Dirac fermions. A nonzero axial density is not consistent with the fermion's mass. Despite its rather exotic definition, the helical chemical potential shares many features with its vector counterpart.

\section{Linear sigma model with quarks}
\label{sec:model}

In order to explore the chiral properties of QCD in the presence of the helical vector potential, we use the linear sigma model coupled to quarks (\M)~\cite{ref:LSMq}. This low-energy effective model of QCD contains two types of fields: the doublet of the light quarks $\psi(x) = (u,d)^T$ and the light pseudoscalars $(\sigma,\vec{\pi})$ which include the pseudoscalar field~$\sigma$ and the isotriplet of the pseudoscalar pions $\vec{\pi} = (\pi_{1},\pi_{2},\pi_{3})$. Each of the light quarks is a triplet in the color space. Since the theory does not contain the gluon (color gauge) fields, the color degeneracy of the quark fields will only lead to the factor $N_f = 3$ in the fermionic contribution to the free energy of the system.

The \M\ Lagrangian has two terms:
\beqn
\cL = \cL_q(\bar \psi, \psi, \sigma, {\vec \pi}, L) + \cL_\sigma(\sigma, \vec \pi) \,,
\label{eq:LL}
\eeqn
The quark part of the Lagrangian~\eq{eq:LL}, 
\beqn
{\cL}_q =
 \overline{\psi} \left[i \slashed\partial - g(\sigma +i\gamma^{5} \vec{\tau} \cdot \vec{\pi})\right]\psi\,,
\label{eq:L:quark}
\eeqn
includes the kinetic term and the interaction between the quark field $\psi$, and the chiral fields $\sigma$ and $\vec \pi$. We do not consider the bare (current) quark mass which is too small to be important for our considerations below. 

The dynamics of the pseudoscalar pions is described by the second term in the Lagrangian~\eq{eq:LL}:
\beqn
{\cL}_{\sigma} (\sigma,\vec{\pi}) & = & \frac{1}{2} \left(\partial_\mu \sigma\partial^\mu \sigma + \partial_\mu \pi^0 \partial^{\mu}\pi^{0}\right) +
\partial_\mu \pi^+ \partial_{\mu}\pi^{-} \nonumber \\
& & - V(\sigma,\vec{\pi})\,,
\label{eq:L:pions}
\eeqn
where we have introduced the fields of the charged and neutral mesons, respectively:
\beqn
\pi^{\pm}=\frac{1}{\sqrt{2}}\left(\pi^{1}\pm i\pi^{2}\right)\,,\qquad \pi^{0}=\pi^{3}\,.
\label{eq:pions}
\eeqn

The potential $V$ in the pionic Lagrangian~\eq{eq:L:pions} contains two terms:
\beqn
V(\sigma, {\vec \pi}) = \frac{\lambda}{4} \left( \sigma^2 + {\vec \pi}^2 - v^2 \right)^2 - h\sigma\,.
\label{eq:V:sigma}
\eeqn 
The first term describes the spontaneous breaking of the chiral symmetry. It leads to a nonzero expectation value of the pseudoscalar field $\avr{\sigma} \neq 0$ and, in general, could also give rise to the emergence of the condensate of pseudoscalar pions $\avr{\vec \pi}$. However, the second term in the same potential \eq{eq:V:sigma} breaks explicitly the symmetry between the components of the pseudoscalar mesons $(\sigma,\vec{\pi})$ and preferentially maximizes the pseudoscalar condensate $\avr{\sigma}$. In addition, this term energetically disfavors the pion condensate: $\avr{\vec \pi} = 0$. As a result, the quarks acquire the dynamical mass $M = g \avr{\sigma}$ via the scalar-quark interaction term of the quark Lagrangian~\eq{eq:L:quark}. 

In our paper, we work in a mean field (MF) approximation thus neglecting quantum fluctuations of the scalar fields $\sigma$ and $\vec{\pi}$. The Lagrangian~\eq{eq:LL} reduces to 
\beqn
\cL_\MF = \overline{\psi} \left(i \slashed\partial - g \sigma \right)\psi - V(\sigma)\,,
\label{eq:LL:MF}
\eeqn
where we take the advantage of the mean-field approximation to simplify the notations for the potential~\eq{eq:V:sigma}, $V(\sigma) \equiv V(\sigma, {\vec 0})$, and for the mean field $\sigma \equiv \avr{\sigma}$. In the mean-field theory~\eq{eq:LL:MF}, the integral over the Dirac fields is taken exactly. 

Following Ref.~\cite{ref:Scavenius}, we adopt the following phenomenological parameters of the model: 
\beqn
g & = & 3.3, \qquad \lambda = 20, \qquad 
v = 87.7\,\mbox{MeV},\nonumber\\
h & = & (114.3\,\mbox{MeV})^3.
\eeqn
With these parameters, the vacuum expectation value of the pseudoscalar field is fixed to the pion decay constant, $\avr{\sigma} = f_\pi = 92.2\, \mbox{MeV}$, the dynamical quark mass 
\beqn
M = g \avr{\sigma}\,,
\label{eq:dynamical:M}
\eeqn
gives us the expected one-third of the mass of a nucleon, $M = 290 \, \mbox{MeV}$, while the tree-level pion mass $m_\pi = \sqrt{\lambda(\avr{\sigma}^2 - v^2)} = 134 \, \mbox{MeV}$ falls in the range of physical pion masses.

\section{Phase structure}

\subsection{Thermodynamics of the sigma model}

We start our investigation of the effects of finite helical density with the phase diagram at vanishing temperature $T = 0$ and then continue to explore the effects of finite helical density on dense quark matter at finite temperature. We consider the dense matter in the plane of the baryonic ($\mu_B$) and helical ($\mu_H$) chemical potentials. The helical chemical potential has been defined earlier, for example, in Eq.~\eq{eq:L:free:VH}. The baryonic chemical potential is taken according to the standard prescription: $\mu_B = N_c \mu_V$, where the vector chemical potential is equal to the quark chemical potential $\mu_V \equiv \mu_q$ and
$N_c = 3$ is the number of colors (three colored quarks constitute one colorless nucleon). 

The fermionic part of the \M\ Lagrangian in the mean-field approximation \eq{eq:LL:MF} is captured by the free-field Lagrangian~\eq{eq:L:free:VH}. The full thermodynamic potential of the model contains the pure pion contribution, given by the potential $V(\sigma)$ and fermionic part, respectively:
\beqn
\Omega\left(\sigma;\mu_V,\mu_H\right) & = & V(\sigma) + \Omega_{q}(\sigma;\mu_V,\mu_H).
\label{eq:Omega:q:T:2}
\eeqn
The fermionic free energy is the sum
\beqn
\Omega_{q}(\sigma;\mu_V,\mu_H) =
\Omega_{\mathrm{vac}}(\sigma)
+ \Omega_T(\sigma;\mu_V,\mu_H),
\label{eq:Omega:q}
\eeqn
of the zero-point (vacuum) part,
\beqn
\Omega_{\mathrm{vac}}(\sigma) = - 12 \int\frac{d^{3}p}{(2\pi)^{3}} \omega_{\p}(\sigma)
\label{eq:Omega:vac:sigma}
\eeqn
and the thermodynamic contribution 
\beqn
& & \Omega_T(\sigma;\mu_V,\mu_H) = - 6 T \sum_{s = \pm 1} \sum_{\varkappa = \pm 1}\int \frac{d^3 p}{(2\pi)^3}  
\label{eq:Omega:T:L0} \\
& & \hskip 7mm \cdot \ln \left(1 + \exp\left\{- \frac{\omega_\p(\sigma) - 
s (\mu_V + \varkappa \mu_H)
}{T}\right\} \right).\quad\
\nonumber
\eeqn
The energy dispersion of the fermions depends on the value of the $\sigma$ condensate:
\beqn
\omega_{\p}(\sigma) = \sqrt{\p^2 + g^2 \sigma^2}.
\label{eq:omega:sigma}
\eeqn

Notice that both matter and temperature influence, via the thermodynamic part~\eq{eq:Omega:T:L0}, the value of the condensate~$\sigma$. The latter quantity determines the fermionic spectrum~\eq{eq:omega:sigma}, which, in turn, appears in the vacuum part of the free energy~\eq{eq:Omega:vac:sigma}. Thus, the matter and temperature effects may modify the value of the vacuum energy in an indirect way.\footnote{Notice that the chirally (axially) imbalanced matter with $\mu_A$ modified the vacuum energy in a direct way~\cite{ref:Marco} thus leading to an explicit divergence~\eq{eq:nA:div}.} This dependence, taken separately, does not pose a problem. However, the vacuum energy needs a regularization with a fixed ultraviolet cutoff, which inevitably enters the effective potential and makes the condensate dependent on the ultraviolet energy scale. As the condensate enters various dimensional physical quantities, the mentioned arbitrariness of the ultraviolet cutoff undervalues the predictive power of the model. Therefore, the ultraviolet-divergent vacuum energy is customarily ignored in thermodynamic studies~\cite{ref:Scavenius,ref:vacuum:logs1} which is justified by the infrared nature of the effective model. We too will ignore the vacuum energy in our approach. 

Taking Eq.~\eq{eq:Omega:T:L0} by parts, we get the more convenient expression for the thermal part of the free energy:
\beqn
& & \Omega_T(\sigma; \mu_V,\mu_H) = - \frac{1}{\pi^2} \sum_{s = \pm 1} \sum_{\varkappa = \pm 1}\int_0^\infty \frac{p^4 d p}{\omega_\p(\sigma)}  
\label{eq:Omega:T:parts} \\
& & \hskip 10mm \times \left(\exp\left\{ \frac{\omega_\p(\sigma) - 
s(\mu_V + \varkappa \mu_H)}{T}\right\} + 1\right)^{-1}.\quad\
\nonumber
\eeqn

The ground state of the model is given by the condensate $\sigma$ which is defined via the minimization of the thermodynamic energy:
\beqn
\Omega\left(\sigma;\mu_V,\mu_H\right) = V(\sigma) + \Omega_{T}(\sigma;\mu_V,\mu_H).
\label{eq:Omega:q:T}
\eeqn
The minimization can generally be carried out numerically. Once the value of $\sigma$ is known, it allows us to find the dynamical mass~\eq{eq:dynamical:M} and determine the phase of the theory. 

We finish this section by providing the expressions for the vector and helical densities which may be obtained from Eq.~\eq{eq:Omega:q} with the help of Eq.~\eq{eq:Q:ell}:
\beqn
n_V & = & \frac{3}{\pi^2} \sum_{s = \pm 1} \sum_{\varkappa = \pm 1} s \int_0^\infty
p^2 d p \nonumber \\
& & \times {\left(\exp\left\{\frac{\omega_\p(\sigma) - 
s(\mu_V + \varkappa \mu_H)}{T}\right\} + 1\right)}^{-1},
\label{eq:nV} \\
n_H & = & \frac{3}{\pi^2} \sum_{s = \pm 1} \sum_{\varkappa = \pm 1} \varkappa \int_0^\infty
p^2 d p \nonumber \\
& & \times {\left(\exp\left\{\frac{\omega_\p(\sigma) - 
s(\mu_V + \varkappa \mu_H)}{T}\right\} + 1\right)}^{-1}.
\label{eq:nH}
\eeqn

\subsection{Dense matter at zero temperature}

In the zero-temperature limit, $T\to 0$, the free energy~\eq{eq:Omega:T:L0} reduces to a simpler form:
\beqn
\Omega_T(\sigma; \mu_V,\mu_H) & = & - \frac{1}{\pi^2} \sum_{s = \pm 1} \sum_{\varkappa = \pm 1}\int_0^\infty \frac{p^4 d p}{\omega_\p(\sigma)}  
\label{eq:Omega:T0} \\
& & \times 
\theta\Bigr[s(\mu_V + \varkappa \mu_H) - \omega_\p(\sigma)\Bigl]
\nonumber
\eeqn
The integral in Eq.~\eq{eq:Omega:T0} can be performed analytically with the help of the identity
\beqn
\int\limits_0^q \frac{p^4 d p}{\varepsilon_p} = \frac{1}{8} \left[q \varepsilon_q (2 q^2 - 3 m^2) + 3 m^4 \arctan
\frac{q}{\varepsilon_q} \right],\quad
\label{eq:integral}
\eeqn
where we denoted $\varepsilon_p = \sqrt{p^2 + m^2}$. Equations~\eq{eq:Omega:T0} and \eq{eq:integral} simplify the numerical calculations.

\begin{figure}[!htb]
\begin{center}
\includegraphics[width=80mm,clip=true]{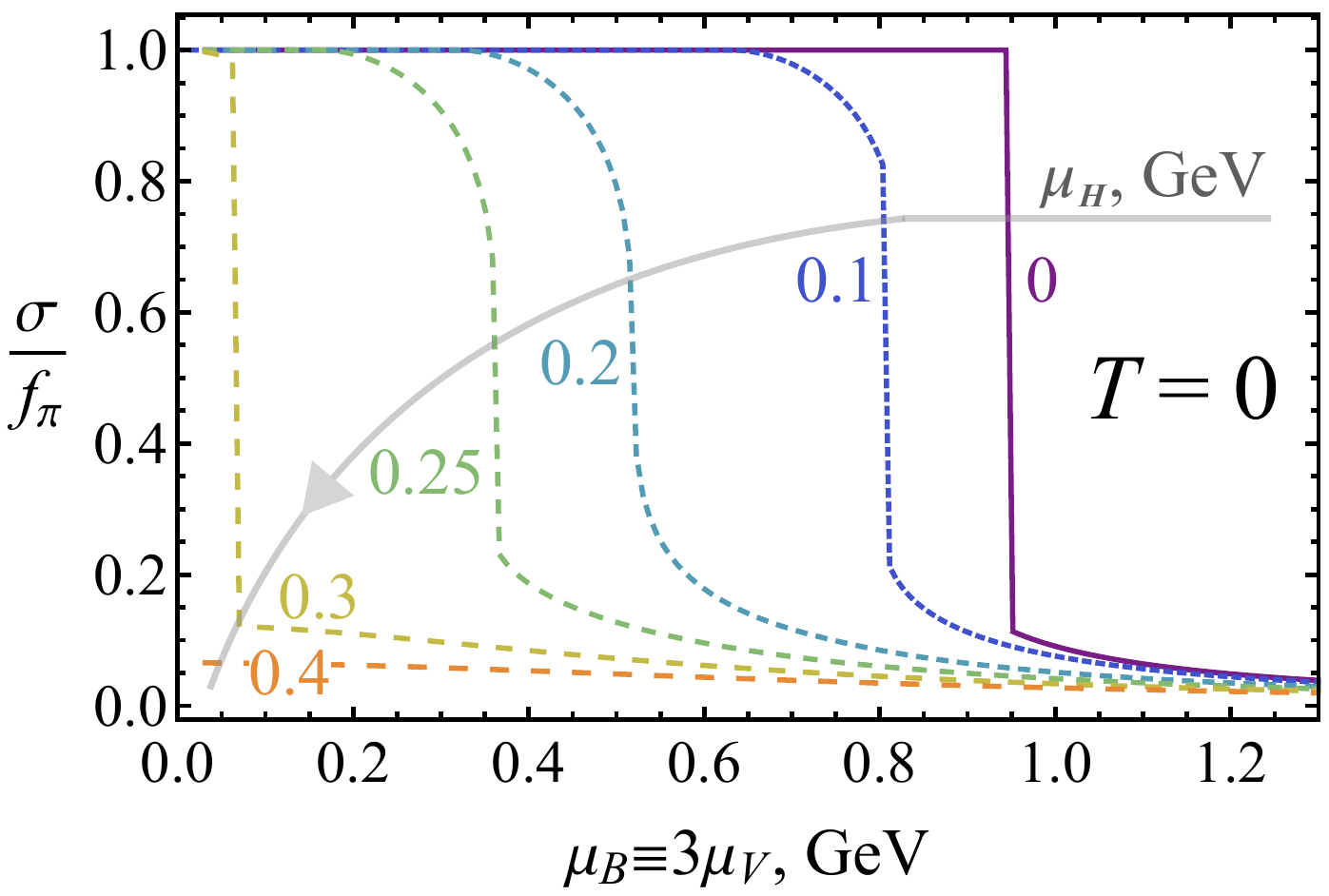} 
\end{center}
\caption{The chiral order parameter $\sigma$ (in units of the pion decay constant in vacuum, $f_\pi$) as a function of the baryonic chemical potential $\mu_B \equiv 3 \mu_V$ at various values of the helical chemical potential $\mu_H$ at zero temperature.}
\label{fig:sigma:T0}
\end{figure}

\begin{figure}[!htb]
\begin{center}
\includegraphics[width=78mm,clip=true]{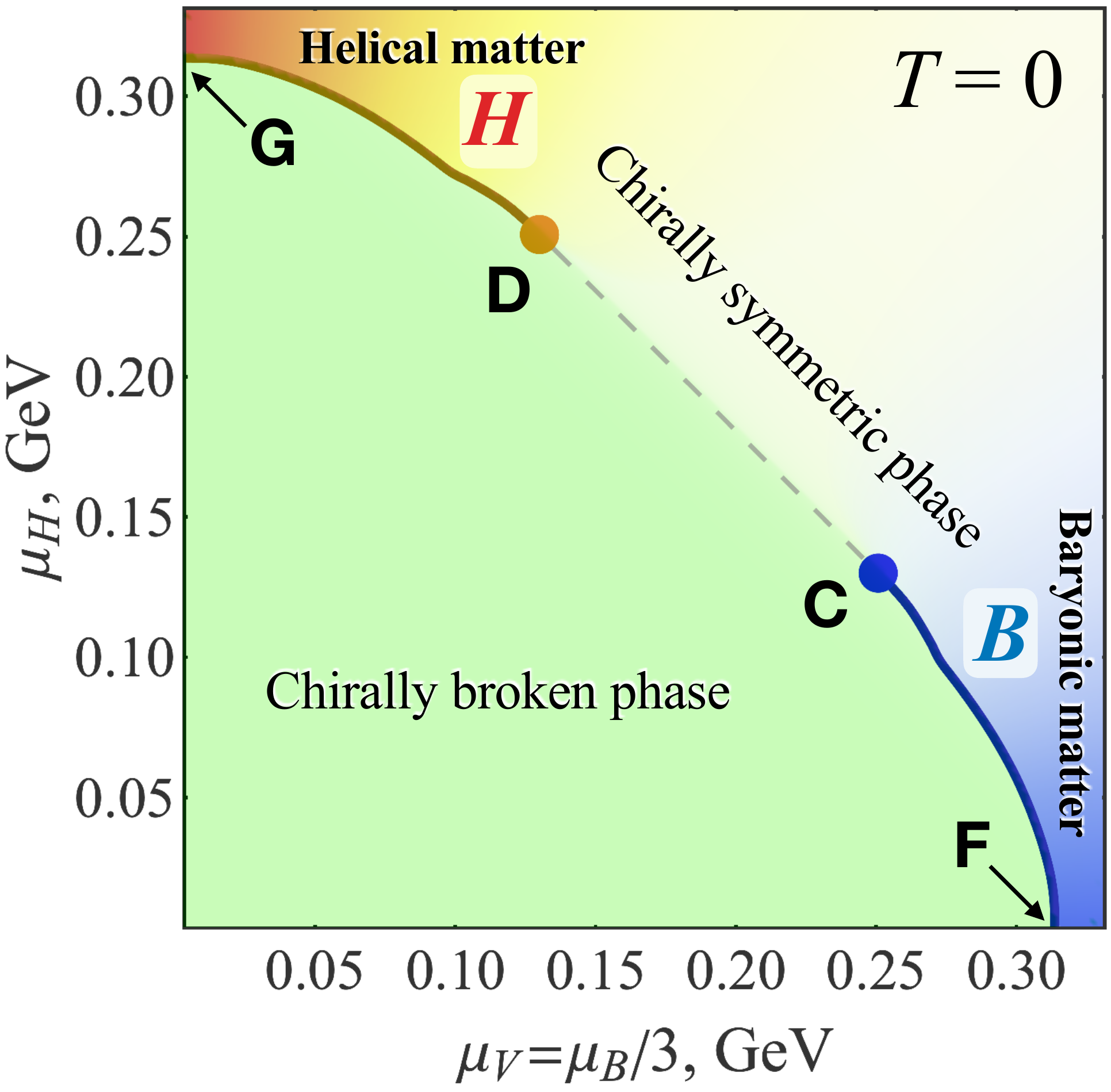} 
\end{center}
\vskip -3mm
\caption{The phase diagram in the ($\mu_V$-$\mu_H$) plane at zero temperature ($\mu_V \equiv \mu_q$). The thick lines mark the first-order phase transitions between the chirally broken phase (at low $\mu_V$ and $\mu_H$) and predominately baryonic (blue) and predominately helical (red) regions. The positions of the points ``F'' and ``G'' are given in Eqs.~\eq{eq:point:F} and \eq{eq:point:G}, respectively. The filled circles ``C'' and ``D'' are the endpoints with the second-order phase transitions [Eqs.~\eq{eq:endpoint:C} and \eq{eq:endpoint:D}]. The dashed line ``C-D'' shows the position of a smooth crossover which connects the endpoints.}
\label{fig:phase:T0}
\end{figure}

In Fig.~\ref{fig:sigma:T0} we show the behavior of the order parameter $\sigma$ as the function of the baryonic chemical potential $\mu_B$ at various values of the helical chemical potential $\mu_H$. At zero helical density, $\mu_H = 0$, the model resides in the chirally broken phase at low baryon densities with $\mu_V < \mu_c$ with 
\beqn
F: \quad 
\left(\begin{array}{c}
\mu_V \\[1mm]
\mu_H  \\[1mm]
T
\end{array}\right)
=
\left(\begin{array}{c}
\mu_c
\\[1mm]
0
 \\[1mm]
 0
 \end{array}\right)
\simeq
\left(\begin{array}{c}
314 \MeV
 \\[1mm]
0
\\[1mm]
0
 \end{array}\right), 
\label{eq:point:F}
\label{eq:mu:c}
\eeqn
where $\mu_c \simeq 314 \MeV$ denotes the critical value of the (vector) chemical potential in \M. The letter ``F'' in Eq.~\eq{eq:point:F} marks the point at the $T=0$ phase diagram in the ($\mu_V, \mu_H$) plane,  Fig.~\ref{fig:phase:T0}, which will be discussed later.

As the helical density increases, the position of the chiral phase transition shifts towards smaller values of the baryon chemical potential $\mu_B$. At the same time, the presence of the helical density softens the transition. Both these effects are seen in  Fig.~\ref{fig:sigma:T0} at the helical chemical potential $\mu_H = 100\MeV$.

A further increase of the helical chemical potential moves the transition to even smaller values of the baryon chemical potential and leads to the disappearance of the first order phase transition which is replaced by
a smooth crossover. These features are seen at $\mu_H = 200\MeV$. 

At higher helical densities, the transition start to strengthen and turns again to a first order phase transition (examples are at $\mu_H = 250\MeV$ and $\mu_H = 300\MeV$ in  Fig.~\ref{fig:sigma:T0}). Finally, as $\mu_H$ is increased, the transition point reaches the lowest possible value at $\mu_V = 0$ and the first-order phase transition disappears altogether. In agreement with the mentioned duality between the vector and helical sectors of the theory, the critical helical potential is tightly related to its vector (baryonic) counterpart~\eq{eq:point:F}:
\beqn
G: \quad 
\left(\begin{array}{c}
\mu_V \\[1mm]
\mu_H  \\[1mm]
T
\end{array}\right)
=
\left(\begin{array}{c}
0
\\[1mm]
\mu_c
 \\[1mm]
 0
 \end{array}\right)
\simeq
\left(\begin{array}{c}
0
\\[1mm]
314 \MeV
 \\[1mm]
 0
 \end{array}\right), 
\label{eq:point:G}
\eeqn
where $\mu_c$ is given in Eq.~\eq{eq:mu:c}.
At a larger helical density, $\mu_H > \mu_{H,c}$, the system resides in the chirally restored phase.  The point ``G'' introduced in Eq.~\eqref{eq:point:G} is also highlighted in Fig.~\ref{fig:phase:T0}.

\begin{figure}[!htb]
\begin{center}
\includegraphics[width=80mm,clip=true]{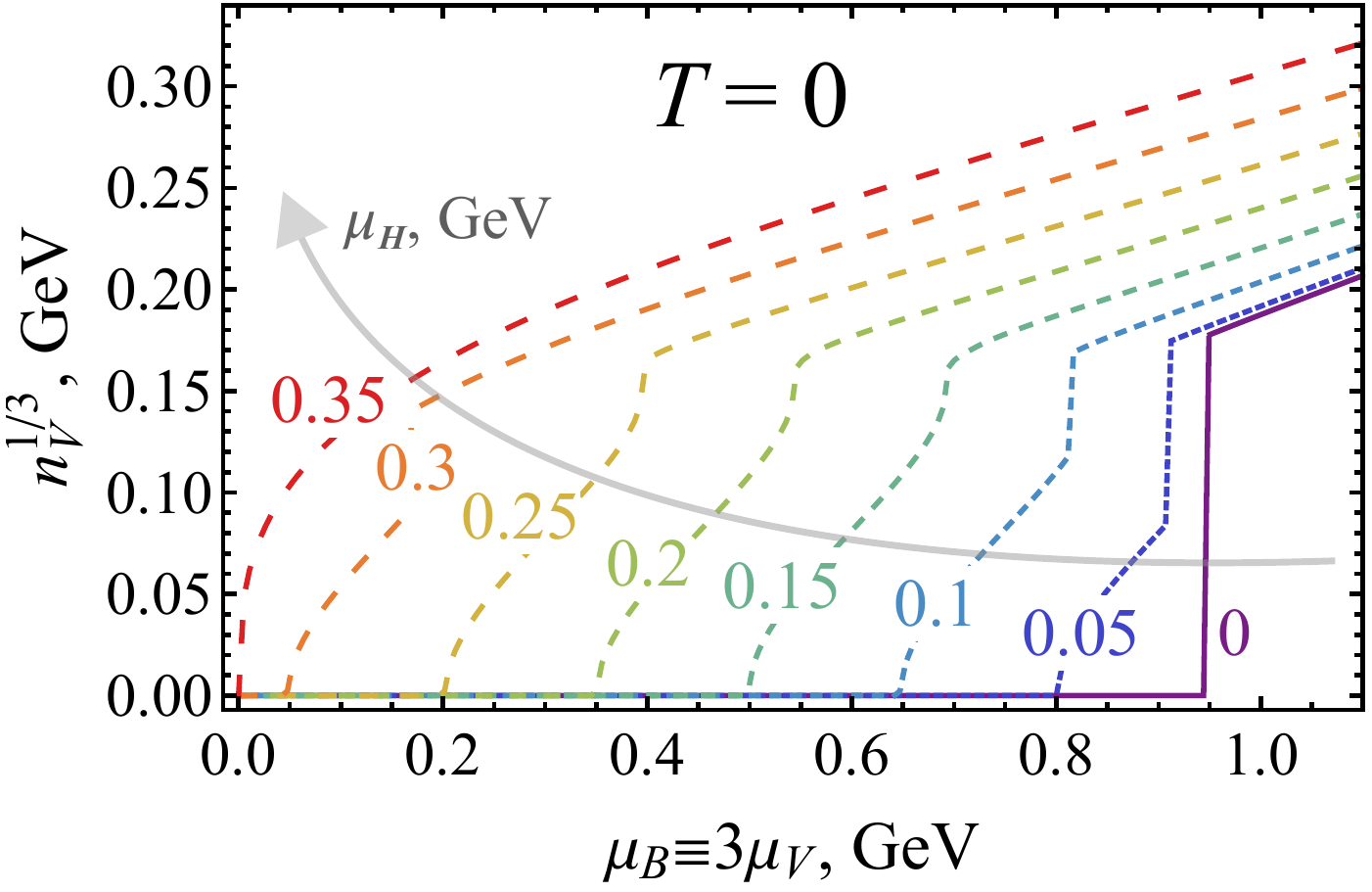} \\[5mm]
\includegraphics[width=80mm,clip=true]{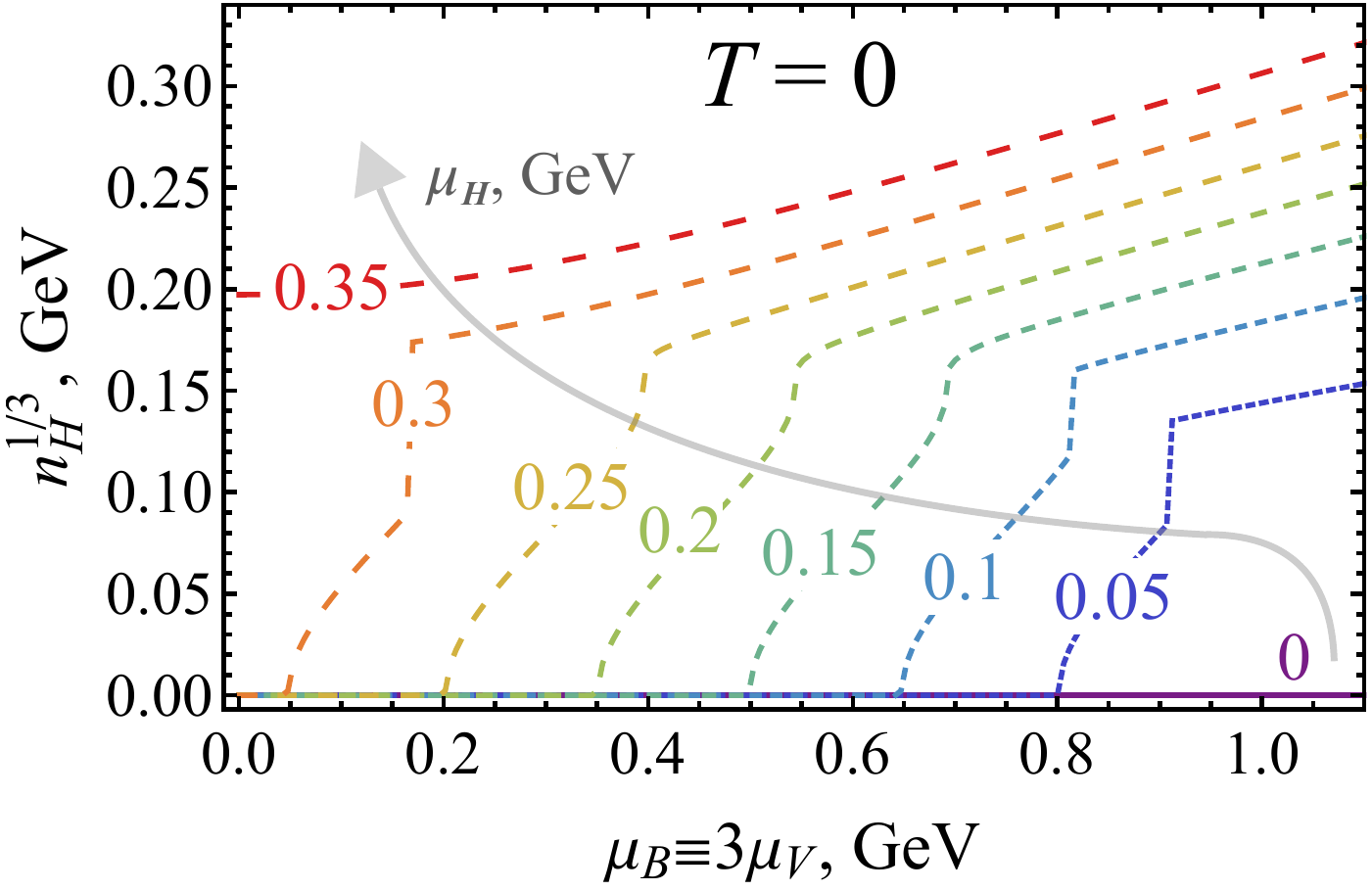} 
\end{center}
\caption{The density of (top) the vector charge $n_V$ and (bottom) the helical charge $n_V = 3 n_B$ as the function of the baryonic chemical potential $\mu_B \equiv \mu_V/3$ at various values of the helical chemical potential $\mu_H$ at zero temperature.}
\label{fig:dens:T0}
\end{figure}

It is instructive to discuss the behavior of the densities of the vector charge\footnote{Due to the relation of the baryon and vector (quark) chemical potentials, $\mu_B = 3 \mu_V = 3 \mu_q$, the baryon density $n_B$ is proportional to the vector (quark) charge density, $n_B = n_V/3 = n_q/3$.}~\eq{eq:nV} and the helical charge~\eq{eq:nH}, shown in Fig.~\eq{fig:dens:T0}. 

At small helical potentials $\mu_H \ll \mu_c$, the first-order chiral phase transition is characterized by a large increase of the baryon density and a small (vanishing at $\mu_H=0$) change in the helical density. 

At moderate values of the helical chemical potential, $\mu_H \sim \mu_c/2$, the chiral crossover transition appears. It is characterized by a smooth change in both vector and helical densities. 

The picture reverses at high values of the helical chemical potential, $\mu_H \sim \mu_c$, where the chiral transition disappears and the helical density prevails over the baryonic (vector) density.

The phase diagram in the plane of chemical potentials  ($\mu_V$, $\mu_H$) is shown in Fig.~\eq{fig:phase:T0}. There are two separate segments of the first-order phase transitions. 

The baryonic segment of the first-order transition begins at the point ``F'' at the zero-helical-density axis~\eq{eq:point:F} and ends at the endpoint ``C'' with the parameters:
\beqn
C: \quad 
\left(\begin{array}{c}
\mu_V \\[1mm]
\mu_H  \\[1mm]
T
\end{array}\right)
\simeq
\left(\begin{array}{c}
242 \MeV \\[1mm]
125 \MeV \\[1mm]
0 
\end{array}\right)
\simeq
\left(\begin{array}{c}
0.77 \mu_c \\[1mm]
0.40 \mu_c \\[1mm]
0
\end{array}\right).
\label{eq:endpoint:C}
\eeqn
This segment separates the chirally broken phase (the green region) from the chirally restored region ``$B$'' where the vector (baryonic) density dominates over the helical charge density. 

The helical segment begins at the point ``G'' at the zero-baryon-density axis~\eq{eq:point:G} and ends at the endpoint ``D'':
\beqn
D: \quad 
\left(\begin{array}{c}
\mu_V \\[1mm]
\mu_H \\[1mm]
T
\end{array}\right)
\simeq
\left(\begin{array}{c}
125 \MeV \\[1mm]
242 \MeV \\[1mm]
0
\end{array}\right)
\simeq
\left(\begin{array}{c}
0.40 \mu_c \\[1mm]
0.77 \mu_c \\[1mm]
0
\end{array}\right).
\label{eq:endpoint:D}
\eeqn
The G-D segment separates the chirally broken phase (the green region) from the chirally restored region ``$H$'' where the helical charge density dominates over the baryonic density. 

The pair of the points ``F'' and ``G'' as well as the pair of the endpoints ``C'' and ``D'' are related to each other by the vector-helical duality~\eq{eq:duality}.

At the endpoints ``C'' and ``D'' a second-order phase transition takes place. These endpoints are connected to each other by a smooth crossover (the dashed thin line ``C-D'').

The phase diagram in Fig.~\eq{fig:dens:T0} is insensitive to the signs of the chemical potentials, being invariant under the separate flips $\mu_V \to - \mu_V$ and $\mu_H \to - \mu_H$.

\subsection{Finite-temperature phase diagram}
\label{sec:phase}

\begin{figure}[!htb]
\begin{center}
\includegraphics[width=85mm,clip=true]{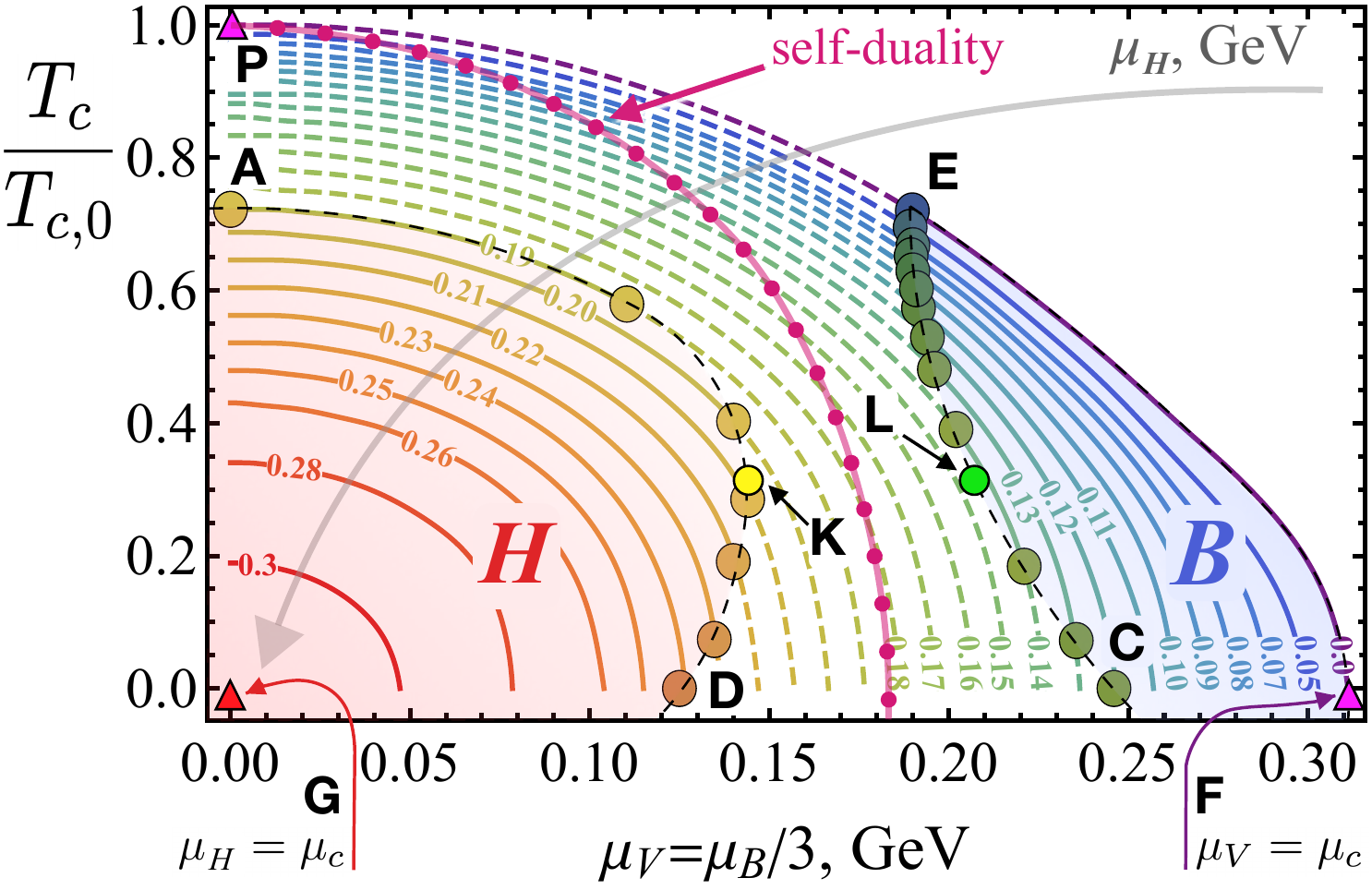} 
\end{center}
\vskip -3mm
\caption{The phase diagram in the ($\mu_V$-$T$) plane at various helical chemical potentials $\mu_H$ (the values of $\mu_H$ in units of GeV are marked at each line). The thick lines mark the first-order phase transitions between the chirally broken phase (the region closer to the origin) and the chirally restored phase (the outer region). The dashed lines show the position of the smooth crossovers. The endpoints of the first-order phase transitions, denoted by the filled circles, are the second-order phase transitions. The segments of the lines of the first-order phase transitions form two ``islands'', with mostly baryonic (blue) and mostly helical (red) chirally-restored phases. The self-duality line $\mu_V=\mu_H$ is shown in the pink color. The meaning of all marked points is discussed in the text.}
\label{fig:phase:T}
\end{figure}

The presence of the helical density leads to a substantial modification of the finite-temperature phase diagram of QCD. In Fig.~\ref{fig:phase:T}, we show the position of the chiral phase transition in the $\mu_V-T$ plane at a dense grid of values of the helical chemical potential $\mu_H$. The temperature is plotted in units of the pseudocritical temperature $T_{c,0}$ of the chiral crossover in \M\ at zero density:
\beqn
T_{c,0} \simeq 144.5 \MeV \qquad  (\mbox{in \M\ at\ } \mu_B = \mu_H = 0). \quad
\label{eq:T:c}
\eeqn
This point is shown by the triangle ``P'' in Fig.~\ref{fig:phase:T}. 

Notice that the exact position of the crossover transition -- which is not associated with a thermodynamic singularity -- depends on the quantity which is used to reveal the crossover. We determine the position of the crossover as a sub-manifold of the parameter space at which the slope of the condensate -- in the $\mu_V-T$ plane at fixed $\mu_H$ -- reaches its maximum. Due to imprecise notion of the position of the crossover, it is customary to call the temperature of the crossover as the ``pseudocritical'' rather than critical temperature.

The value of the critical temperature~\eq{eq:T:c} in \M\ is slightly (less than $10\%$) lower than the value of the critical temperature $T_c^{\mathrm{QCD}} = 156.5(1.5)\,\MeV$ of the chiral phase transition determined via the inflection point of the light-quark chiral condensate in the first-principle simulations of lattice QCD with real quark masses~\cite{Bazavov:2018mes}.

At a vanishing helical chemical potential, $\mu_H = 0$, we recover the standard chiral transition which is plotted at the outer violet line in Fig.~\ref{fig:phase:T}. At small values of the chemical potential $\mu_V$, the chiral transition is a smooth thermodynamic crossover (the dashed line), which reaches the zero-density axis at the point ``P'' with $T=T_{c,0}$ and $\mu_V = \mu_H = 0$. 

As the $\mu_V \equiv \mu_B/3$ chemical potential increases, the crossover turns into the 1st order phase transition (the solid line) passing via the 2nd order endpoint ``E'' (the filled circle). This endpoint is located at:
\beqn
E: \quad 
\left(\begin{array}{c}
\mu_V \\[1mm]
\mu_H \\[1mm]
T
\end{array}\right)
\simeq
\left(\begin{array}{c}
190 \MeV \\[1mm]
0 \\[1mm]
104 \MeV
\end{array}\right) 
\simeq
\left(\begin{array}{c}
0.61 \mu_c \\[1mm]
0 \\[1mm]
0.72 T_{c,0}
\end{array}\right).
\label{eq:endpoint:E}
\eeqn
At higher baryon densities, the first-order transition line segment hits the $T=0$ axis at the point ``F'', given in Eq.~\eq{eq:point:F}. At even higher $\mu_V$, the model resides in the chirally restored phase.

Figure~\ref{fig:phase:T} demonstrates that the presence of a non-zero helical density changes the chiral phase transition substantially. Let us consider what happens with the chiral phase transition as the helical density increases.

First, as the helical chemical potential $\mu_H$ raises, the line of the chiral transition gradually shrinks towards the origin $(\mu_V,T) = (0,0)$. In other words, both the critical temperature $T_c$ at a fixed chemical potential $\mu_B$ and the critical chemical potential $\mu_B$ at fixed temperature $T$ are monotonically decreasing functions of the helical chemical potential $\mu_H$. 

Second, the presence of a moderate helical density makes the chiral phase transition weaker: the line of the first order phase transition shrinks, in favor of the smooth crossover. As $\mu_H$ is increased from $0$ up to $\mu_H = 125 \MeV$, the high-temperature endpoint $E$ descends towards smaller temperatures and larger values of $\mu_V$.

When the helical chemical potential reaches the value $\mu_H = 125 \MeV$, the lower end of the 1st order line reaches the endpoint ``C'' of the $T=0$ phase diagram with $\mu_V = 242 \MeV$. The position of this endpoint, given explicitly in Eq.~\eq{eq:endpoint:C}, is shown in Fig.~\ref{fig:phase:T0}. 

As soon as the helical potential becomes larger than $\mu_H = 125 \MeV$, the chiral phase transition develops the second endpoint, now at the low-temperature end of the 1st order line. The rest of the transition line is occupied by the smooth crossover all the way down to the $T=0$ axis. The structure of the chiral transition line with two endpoints is clearly seen in Fig.~\ref{fig:phase:T} for values of $\mu_H$ between $\mu_H(E) = 125 \MeV$ and $\mu_H(L) = 142 \MeV$, where the point ``L'' is described below.

Further increasing $\mu_H$ causes both the higher- and the lower-temperature endpoints to approach each other. This effect is seen at the line with the fixed value of the chiral helical potential $\mu_H = 140 \MeV$. At a higher, critical value of $\mu_H$, the first-order segment shrinks to zero and the chiral transition turns into a smooth crossover. The 1st order phase transition disappears at the point ``L'' with the parameters
\beqn
L: \quad 
\left(\begin{array}{c}
\mu_V \\[1mm]
\mu_H \\[1mm]
T 
\end{array}\right)
\simeq
\left(\begin{array}{c}
208 \MeV \\[1mm]
142 \MeV \\[1mm]
46 \MeV
\end{array}\right)
\simeq
\left(\begin{array}{c}
0.66 \mu_c \\[1mm]
0.45 \mu_c \\[1mm]
0.32 T_{c,0}
\end{array}\right).
\label{eq:endpoint:L}
\eeqn
which is shown in Fig.~\ref{fig:phase:T} as a bright green dot. 

After the system goes beyond the point ``L'', the chiral phase transition keeps the crossover type for a while until the helical chemical reaches the value $\mu_H \simeq 208 \MeV$. The first-order phase transition reappears at the point ``K'' of the phase diagram:
\beqn
K: \quad 
\left(\begin{array}{c}
\mu_V \\[1mm]
\mu_H \\[1mm]
T 
\end{array}\right)
\simeq
\left(\begin{array}{c}
142 \MeV \\[1mm]
208 \MeV \\[1mm]
46 \MeV
\end{array}\right)
\simeq
\left(\begin{array}{c}
0.45 \mu_c \\[1mm]
0.66 \mu_c \\[1mm]
0.32 T_{c,0}
\end{array}\right).
\label{eq:endpoint:K}
\eeqn
It's easy to see that the points L and K are related to each other via vector-helical duality~\eq{eq:duality}.

In the narrow range of values of the helical chiral potential,
\beqn
0.45 \mu_c \simeq 142 \MeV \lesssim \mu_H \lesssim 208 \MeV \simeq 0.66 \mu_c,
\label{ref:mu:H:cross}
\eeqn
the whole line of the chiral transition is of the crossover type. The crossover region of the phase diagram is cut in two parts by the self-duality line, $\mu_V = \mu_H$, shown in the pink color in Fig.~\ref{fig:phase:T}.

As $\mu_H$ increases  beyond the point K ($\mu_H \simeq 208 \MeV$), the first-order phase transition reappears. Contrary to the low-helicity case, now the first order chiral segment is located at higher temperatures, while the crossover transition is now realized at the colder part of the phase diagram. 

The line of the first order phase transition reaches the zero-baryon density $\mu_V=0$ at the point 
\beqn
A: \quad 
\left(\begin{array}{c}
\mu_V \\[1mm]
\mu_H \\[1mm]
T
\end{array}\right)
\simeq
\left(\begin{array}{c}
0 \\[1mm]
190 \MeV \\[1mm]
104 \MeV
\end{array}\right) 
\simeq
\left(\begin{array}{c}
0 \\[1mm]
0.61 \mu_c \\[1mm]
0.72 T_{c,0}
\end{array}\right),
\label{eq:endpoint:A}
\eeqn
which is exactly dual to the endpoint ``E'', Eq.~\eq{eq:endpoint:E}, of the usual standard finite density transition.

When the helical potential reaches the value $\mu_H \simeq 242 \MeV \simeq 0.77 \mu_c$, the crossover shrinks to zero and the whole chiral phase transition becomes of the first order. The chiral transition intersects with the $T=0$ axis at point D with parameters shown in Eq.~\eq{eq:endpoint:D}. This point is also shown in the $T=0$ phase diagram of Fig.~\ref{fig:phase:T0}.

After passing point D, the crossover region disappears completely and the transition line becomes the first--order phase transition. The temperature and the vector chemical potential decreases monotonically as $\mu_H$ increases. The chiral transition disappears altogether when the chiral helical potential reaches the value $\mu_H = \mu_c$ with $\mu_c$ given in Eq.~\eq{eq:mu:c}. This position marks the point G, Eq.~\eq{eq:point:G}, shown in Figs.~\ref{fig:phase:T0} and \ref{fig:phase:T}.

Qualitatively, the phase diagram in the $\mu_V-T$ plane, Fig.~\ref{fig:phase:T}, has three distinct regions, including the strait of the crossover transition at intermediate values of the helical potential~\eq{ref:mu:H:cross} which separates the high-$\mu_V$ island of the baryonic-rich first-order transitions and the low-$\mu_V$ island of the helical-rich first-order transitions (cf. Fig.~\ref{fig:phase:T0}). The crossover strait is cut in two pieces by the self-duality line, $\mu_V = \mu_H$.

\begin{figure*}[!htb]
\begin{center}
\includegraphics[width=160mm,clip=true]{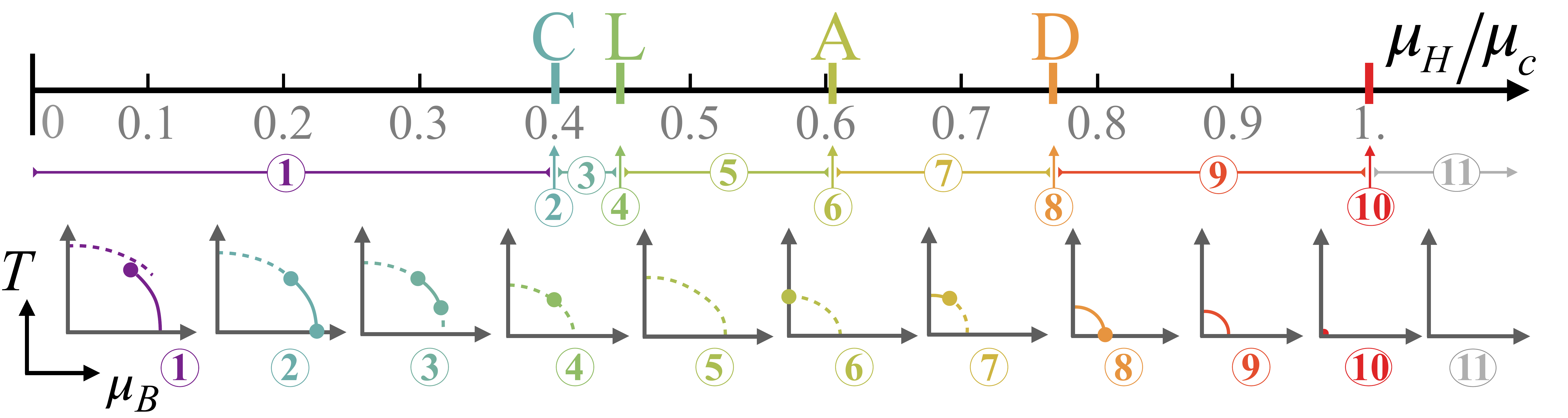} 
\end{center}
\vskip -3mm
\caption{
The evolution of the structure of the chiral transition in QCD as $\mu_H$ (given in units of $\mu_c \simeq 314 \MeV$) is increased. The points C, L, A, and D correspond to the ones given in Fig.~\ref{fig:phase:T}.}
\label{fig:evolution}
\end{figure*}

The evolution of the structure of the chiral transition in the $\mu_V-T$ plane as the helical chemical potential is increased is shown in Fig.~\ref{fig:evolution}. The top row highlights the position on the $\mu_H / \mu_c$ axis of the points $C$, $L$, $A$, and $D$ introduced in Fig.~\ref{fig:phase:T}. The bottom row shows qualitatively the phase diagram in the $\mu_V-T$ plane via $11$ representative configurations at or around these points.

\subsection{Curvature of the chiral transition}

One of the most important characteristics of the chiral  transition is the curvature $\kappa$ of the transition temperature at small values of the baryon chemical potential $\mu_B \equiv 3 \mu_V$:
\beqn
\frac{T_c(\mu_B,\mu_H)}{T_{c,0}} = \frac{T_c(\mu_H)}{T_{c,0}} - \kappa(\mu_H) \left( \frac{\mu_B}{T_{c,0}} \right)^2 + \dots,
\label{eq:curvature:exp}
\eeqn
where $T_c(\mu_H) \equiv T_c(\mu_B = 0, \mu_H)$ is the (pseudo)critical temperature of the chiral transition at zero baryonic density, and $T_{c,0}\equiv T_c(\mu_B = 0, \mu_H = 0)$ is the position of the crossover when both chemical potentials vanish~\eq{eq:T:c}. Due to the vector-helical duality~\eq{eq:duality}, the dependence of the (pseudo)critical temperature $T_c(\mu_H) \equiv T_c(0,\mu_H)$ on helical chemical potential $\mu_H$ at $\mu_V = 0$ is the same as the dependence $T_c(\mu_V) \equiv T_c(3\mu_V,0)$ on $\mu_V$ at $\mu_H = 0$, given by the outer transition line in Fig.~\ref{fig:phase:T}.

The series in Eq.~\eq{eq:curvature:exp} is valid provided the baryon potential is much smaller than the chiral transition temperature at zero density~\eq{eq:T:c}, $\mu_B \ll T_{c,0}$. The linear term is absent due to the $C$-symmetry of the theory at the baryonic neutrality point, $\mu_B = 0$. The dots in the series~\eq{eq:curvature:exp} represent higher-order terms in $\mu_B/T_{c,0}$.

\begin{figure}[!thb]
\begin{center}
\includegraphics[width=85mm,clip=true]{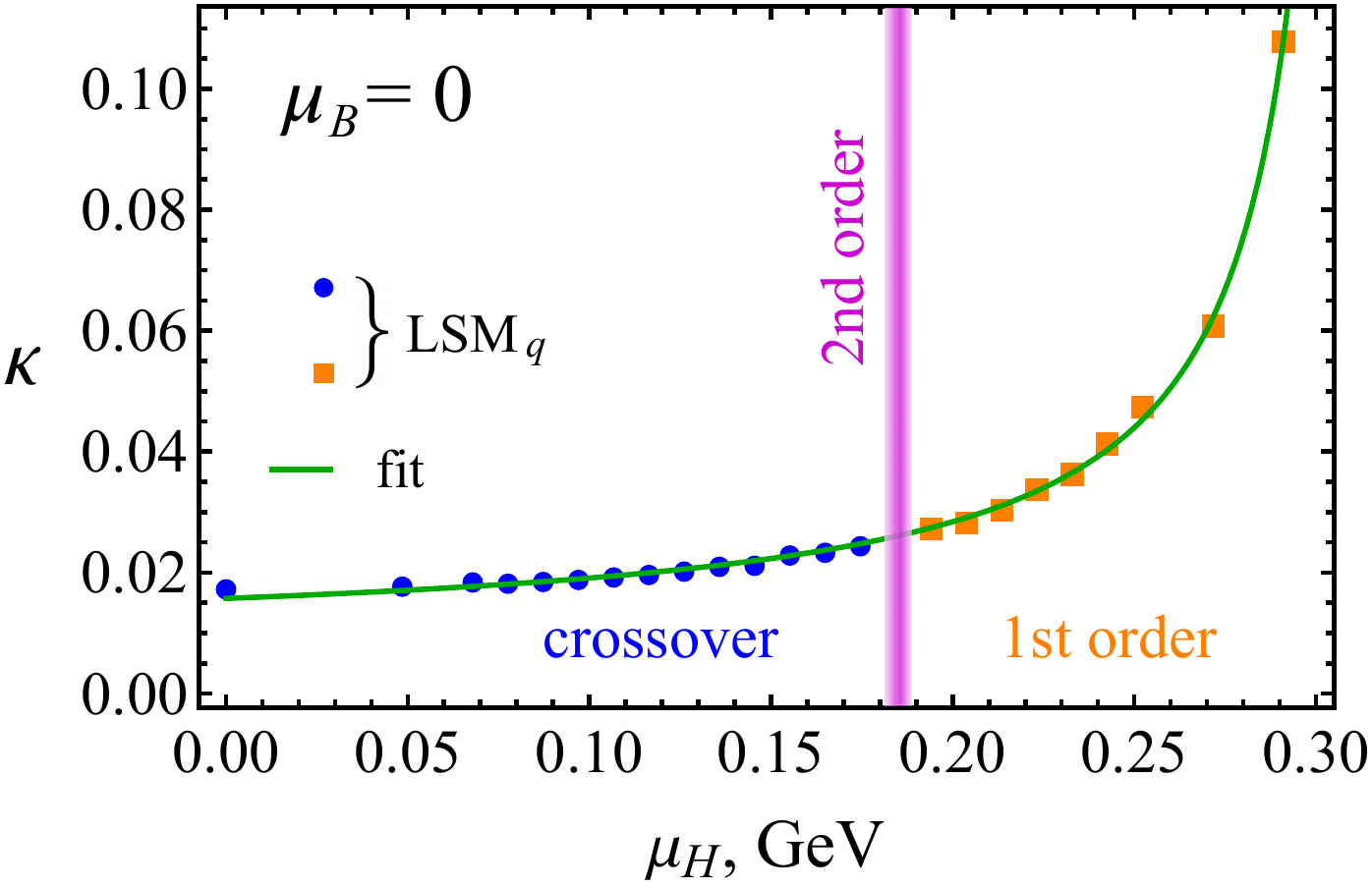} 
\end{center}
\vskip -3mm
\caption{Curvature $\kappa$ of the chiral transition~\eq{eq:curvature:exp} as the function of the helical chemical potential $\mu_H$ at zero baryon density $\mu_B = \mu_V = 0$. As the helical density rises, the crossover (the blue dots) changes into the 1st order phase transition (the orange squares). The position of the 2nd order endpoint (point ``A'' in Fig.~\ref{fig:phase:T}) is shown by the vertical magenta line. The green curve represents the best fit~\eq{eq:fit}.}
\label{fig:curvature}
\end{figure}

It turns out that the first two terms in Eq.~\eq{eq:curvature:exp} may indeed describe very well the curvature of the chiral transition lines, shown in Fig.~\ref{fig:phase:T}, at any helical chemical potential $\mu_H$. The effect of the helical density on the curvature $\kappa$ of the chiral transition is shown in Fig.~\ref{fig:curvature}.

The curvature of the phase transition is a slowly rising function of the helical chemical potential $\mu_H$. After the helical density passes the critical 2nd-order endpoint  ``A'', Eq.~\eq{eq:endpoint:A}, the crossover turns into a first order phase transition with the rising, as a function of $\mu_H$, curvature. The evolution of the curvature slope suggests the presence of a singularity at the point ``G'', Eq.~\eq{eq:point:G}, where the curvature becomes infinite. 

Both crossover and 1st order regions may be successfully described by the same fitting function in the whole region of $\mu_H$:
\beqn
\kappa^{\mathrm{fit}}(\mu_H) = \kappa_0 \left( 1 + \alpha \frac{ 
\mu_H}{\mu_{H,c} - \mu_H} \right).
\label{eq:fit}
\eeqn
The fitting parameters are the value of the curvature at the vanishing helical density, $\kappa_0 \equiv \kappa(\mu_H =0)$, the critical value of the helical chemical potential, $\mu_{H,c}$, and the coefficient $\alpha$ which controls the slope of Eq.~\eq{eq:fit}.

The best fit, shown in Fig.~\ref{fig:curvature} by the green line, gives the following values for the $\mu_H = 0$ curvature and the critical value of the helical chemical potential:
\beqn
\kappa_0 = 0.0158(3), 
\qquad
\mu_{H,c} = 0.314(1)
\label{eq:best:fits}
\eeqn
For the slope, we get $\alpha = 0.46(2) \approx 1/2$. 

The value of the curvature~\eq{eq:best:fits} at zero helical density, predicted by \M, is well-compatible with the QCD result $\kappa^{\mathrm{QCD}} = 0.0132(18)$ of the pseudocritical line of the chiral transition obtained with the help of first-principle lattice simulations~\cite{Bonati:2014rfa}.

An increasing helical density leads to the rise of the curvature $\kappa$. The curvature becomes infinite at the critical value $\mu_{H,c}$ of the helical chemical potential~\eq{eq:best:fits} which expectedly coincides with the point ``G'' within the errors~\eq{eq:point:G}.

A few words on the validity of our results are in order. As the \M\ is an effective low-energy (infrared) model of QCD, at large values of massive parameters (for example, at high chemical potentials) the model may give somewhat inaccurate results. Therefore, as a standard word of caution, our quantitative predictions should be considered with certain care.

In addition, following the traditional approach, we characterized the vacuum of the theory with a single parameter: the pseudoscalar condensate $\sigma$. This single condensate can indeed describe the vacuum at low baryon densities both in the chirally broken phase and around the chiral transition. Due to the vector-helical duality, the single pseudoscalar condensate may be used in the similar range of values of the helical chemical potential $\mu_H$ (at low helical densities). However, the presence of large helical and baryonic chemical potentials -- for example, around the crossover transition -- may lead to the formation of new types of condensates which may not be reduced to the single parameter $\sigma$.

\section{Discussion and Conclusion}
\label{sec:conclusions}

\subsection{Thermodynamic relevance of helicity}

In our paper we discussed the influence of the presence of a finite helical density on the phase diagram of QCD at finite temperature and finite baryon density. 

For quarks, a difference between the helicity and chirality appears at the level of their transformations under the charge conjugation operation ($C$).\footnote{The term ``axiality'', which would be a more appropriate term than ``chirality'', is not adopted in the current QCD literature.} The right- (left-)handed helicity corresponds to a positive (negative) value of the projection of the quark's spin on its momentum.  The chirality of the quark is given by its helicity times the sign of the particle's charge: the chirality of a particle is equal to its helicity (for example, a right-chiral particle has a right-handed helicity) while the chirality of an antiparticle is opposite to its helicity (for instance, a right-chiral antiparticle has a left-handed helicity). 

Physically, the notions of helicity and chirality are very close to each other as they differ only by an application of the vector charge operator, $Q_V$. However, the helicity can be defined for free massive quarks while the chirality cannot. Indeed, the helicity operator $h$, given in Eq.~\eq{eq:h}, commutes with the Hamiltonian of a massive free fermion~\eq{eq:h:H} while the axial operator, given by the $\gamma^5$ matrix, does not. 

Mathematically, the axial charge density is determined via the local operator $\gamma^5$, while the helical density involves a more complicated expression~\eq{eq:h}. Moreover, the definition of the axial charge is a Lorentz invariant while the helical density depends on the choice of the local reference frame. The latter property, however, is not important for systems at finite density and temperature (as the one considered in this paper), where the Lorentz symmetry is explicitly broken by the presence of matter.

We argue that the helical density is a thermodynamically relevant quantity in theories with the mass gap generation such as QCD. For example, both the helical charge and the helical chemical potential $\mu_H$ are well-defined quantities in a thermodynamic ensemble of free massive fermions. On the contrary, the axial symmetry is inconsistent with the massive quarks: the axial chemical potential $\mu_A$ modifies the energy spectrum of the massive fermions and leads to a $\mu_A$-dependent divergence of zero-point fluctuations~\cite{ref:Marco}. The latter property casts a shadow on the very definition of the vacuum of the theory at nonzero $\mu_A$: the zero-point fluctuations are associated with the vacuum and should, therefore, be independent of the presence of matter given by a nonvanishing chemical potential.

It is worth discussing the relevance of the effects of a finite helical density to quark-gluon plasma created in relativistic heavy-ion collisions. The helicity fluctuations are likely to emerge in the initial stages of heavy-ion collisions in an off-equilibrium regime. While the quarks may be created with a net helicity, the magnitude of the helical charge remains yet to be estimated. Nevertheless, the global helicity number is expected to be approximately conserved in the high-temperature phase before
the hadronization stage is reached. Indeed, it is well-known that the helicity of massless quarks is conserved in perturbative QCD interactions due to the vector coupling to gluons (see, for example, the discussions in Refs.~\cite{Kapusta:2019sad,Kapusta:2019ktm,Kapusta:2020npk}). For an ultra-relativistic quark with a mass small compared to its energy, the perturbative helicity-flip cross-section is proportional to the quark mass squared. 

Nonperturbative interactions and U(1) symmetry breaking may increase the helicity flip rate~\cite{Kapusta:2019ktm} (see also Ref.~\cite{Ruggieri:2016asg} where the chirality-flip rate in the QGP was addressed in an effective approach). Therefore it is reasonable to expect that the helicity is good conserved quantity in the chirally-symmetric phase of QCD where the masses of the light quarks are small compared to their thermal energies. 

In our paper, we restrict ourselves to the simplest case of two light quarks $u$ and $d$. It is worth noticing that the helicity flip should occur in interactions with a massive quarks and can become relevant for the thermalization of the spin of a massive $s$-quark with a (rotating, for example) environment. The thermalization can occur -- via a mechanism involving breaking of an axial U(1) symmetry -- shortly before the QGP reaches the hadronization stage~\cite{Kapusta:2019ktm}. For the realistic parameters of the quark-gluon plasma (QGP), the quark's helicity and its spin equilibrate at the same rate~\cite{Kapusta:2020npk}. Therefore, the spin of an $s$-quark may pick up the direction of the local vorticity of the rotating quark-gluon plasma during its evolution after a heavy-ion collision, thus leaving an experimentally observed imprint on polarization of $\Lambda$ hyperons. 

Summarizing, it is generally expected that the quarks are kinetically thermalized within a short time of the order of $0.5\,\mbox{fm}/c$ after the collision. The next $5-10\,\mbox{fm}/c$ until hadronization, the fireball evolution is described by an approximately thermalized QGP~\cite{Heinz:2013ar}. The thermalized light quarks carry the conserved net helicity which should, as we show in our article, affect the QCD phase diagram and thus may influence the evolution of QGP.

\subsection{Effects of net helicity on QGP thermodynamics}

After establishing the consistency of the helical density with the mass gap generation, we studied the influence of the helical chemical potential on the chiral properties of QCD. To this end, we used the linear sigma model coupled to quarks. We demonstrated that the presence of the helical density affects the phase diagram of dense quark matter in a rather complicated way, both at zero (Fig.~\ref{fig:phase:T0}) and finite (Fig.~\ref{fig:phase:T}) temperature. We used for the analysis the vector chemical potential $\mu_V$ which is equivalent to the quark chemical potential $\mu_q$ and three times smaller than the baryon chemical potential, $\mu_B$: $\mu_V \equiv \mu_B/3 \equiv \mu_q$.

The evolution of the structure of the chiral transition in dense QCD as a function of the helical chemical potential is summarized in Fig.~\ref{fig:evolution}. Here we summarize the main effects of the helical chemical potential on the chiral transition:
\begin{enumerate}
    \item A moderate helical density makes the chiral phase transition softer while shifting the critical endpoint towards lower temperatures and higher baryon chemical potentials. 

    \item In a certain narrow range of the helical chemical potential $\mu_H$, the chiral phase transition acquires an additional 2nd order endpoint.

    \item At intermediate helical density, the segment of the first-order transition disappears and the chiral transition becomes a soft crossover at any temperature or baryonic density. 
    
    \item At even higher helical chemical potentials, the first-order transition reappears: the finite-$T$ phase transition at zero baryon density ($\mu_B = 0$) becomes of first order, which turns into crossover at a nonzero $\mu_V$.

    \item Finally, the chiral transition turns into a 1st order transition at any temperature and baryonic density, before disappearing altogether when the helical chemical potential reaches the critical value, $\mu_H = \mu_c$.
\end{enumerate}

We have also demonstrated the existence of a thermodynamic duality between the helical and vector (baryonic) chemical potentials~\eq{eq:duality}: the fermionic free energy is invariant under a permutation of the vector and helical chemical potentials~\eq{eq:VH:symmetry}. This duality should (softly) be broken by the electromagnetic interactions that were not considered in this article.

In relativistic heavy-ion collisions, the quark-gluon plasma is created in a low-density regime characterized by small values of the baryon chemical potential. The region of a low baryon chemical potential is well accessible in the first-principle simulations of lattice QCD. At small baryonic densities, the chiral transition is a smooth crossover with the pseudocritical temperature diminishing quadratically as a function of the baryonic chemical potential~\eq{eq:curvature:exp}. 

In the limit of vanishing global helicity, $\mu_H = 0$, our result for the curvature of the pseudocritical temperature~\eq{eq:best:fits} agrees reasonably well with the results of the lattice simulations~\cite{Bonati:2014rfa}. The presence of a nonvanishing density of helical quark charges enhances the curvature of the (pseudo)critical line of the chiral transition at low baryon density, Fig.~\ref{fig:curvature}. The curvature diverges as the helical chemical potential reaches the critical point, $\mu_H = \mu_c \simeq 314 \MeV$.

\acknowledgments

The authors are grateful to M.~Ruggieri for interesting discussions and useful comments. M.N.C. is partially supported by Grant No. 0657-2020-0015 of the Ministry of Science and Higher Education of Russia. The work of V.E.A. is supported by a Grant from the Romanian  National Authority for Scientific Research and Innovation, CNCS-UEFISCDI, project number PN-III-P1-1.1-PD-2016-1423.

\end{document}